\documentclass[aps,11pt]{revtex4-1}

\newtheorem{theorem}{Theorem}

\newtheorem{lemma}{Lemma}
\newtheorem{corollary}{Corollary}

\usepackage{graphicx}
\usepackage{amsfonts}
\usepackage{amsmath}
\usepackage{mathrsfs}
\usepackage{eucal}
\usepackage{appendix}
\usepackage[ruled]{algorithm2e}
\usepackage{natbib}

\newcommand\bR{\mathbb{R}}

\newcommand\mF{\mathcal{F}}

\newcommand\bC{\mathbf{C}}
\newcommand\bA{\mathbf{A}}
\newcommand\bD{\mathbf{D}}

\def\o{\omega}

\newcommand\qed{$\square$}

\begin{document}



\title{Towards the full information chain theory: answer depth and source models}
\author{E. Perevalov}
\email[E-mail: ]{eup2@lehigh.edu}
\author{D. Grace}
\email[E-mail: ]{dpg3@lehigh.edu}
\affiliation{Lehigh University\\ Bethlehem, PA}
\date{\today}

\begin{abstract}
A problem of optimal information acquisition for its use in general decision making problems is considered. This motivates the need for developing quantitative measures of information sources' capabilities for supplying accurate information depending on the particular content of the latter.
A companion article developed the notion of a question difficulty functional for questions concerning input data for a decision making problem. Here, answers which an information source may provide in response to such questions are considered. In particular, a real valued answer depth functional measuring the degree of accuracy of such answers is introduced and its overall form is derived under the assumption of isotropic knowledge structure of the information source. Additionally, information source models that relate answer depth to question difficulty are discussed. It turns out to be possible to introduce a notion of an information source capacity as the highest value of the answer depth the source is capable of providing.
\end{abstract}



\maketitle

\section{\label{s:intro}Introduction}
\begin{figure}
\includegraphics{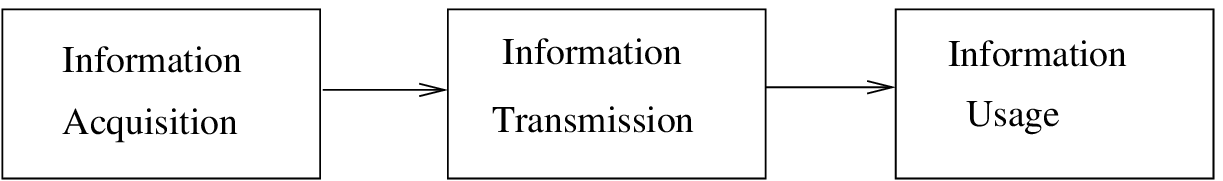}
\caption{\label{f:Ichain}Full information chain}
\end{figure}

The classical Information Theory was developed as a theory of communication, its main practical objective being optimization of communication over imperfect channels. Correspondingly, it deals with information {\it quantity}, while paying little or no attention to either its {\it accuracy} or {\it relevance}. The latter omission is by no means a defect of Information Theory but rather its conscious choice: the complete abstraction from any content of transmitted information ensured both the theory universality and its notable elegance. On the other hand, besides being transmitted, information also gets acquired and used in everyday practice of a variety of fields, including science and engineering. This typical path of information, from acquisition, via possible transmission, to its usage (to make decisions, generate new knowledge etc.) can be schematically depicted as the {\it full information chain} shown in Fig.~\ref{f:Ichain}. One can see, that unlike the middle link of this chain, the two ``end'' links do not at this moment enjoy the convenience of being described by any kind of universal theory. While several methods for making decisions under incomplete information have been developed in considerable detail and used in numerous applications, the overall theory providing a unified and explicit treatment of informational aspect of such activity is still largely lacking. In fact, the state-of-the-art of (broadly defined) decision making under uncertainty could be compared to that of theory and practice of information transmission before the advent of Information Theory in late 40's \cite{SHANNON:1948}. Various coding schemes (like Morse code, for instance) existed and were widely used, but they were developed in largely ``trial and error'' fashion, and, for example, their optimality and theoretical limits were generally unknown.

One of the goals of the present article is to initiate the development of an information theory of the ``end'' links of the information chain shown in Fig.~\ref{f:Ichain}. This would make it necessary, in particular, to explicate the quantitative properties of information accuracy and relevance, in addition to its quantity. This article, which a follow-up to \cite{part1}, is mostly devoted to developing a theory of the first link of the full information chain: the information acquisition link. While a consistent theory of the two ``end'' links of the full information chain appears to require a joint treatment of both links, this article, together with \cite{part1}, develops the basic ``machinery'' of the information acquisition link analysis. It is nevertheless necessary to emphasize that the proposed theory can only be logically complete (even at the most basic level of detail) only when the third (information usage) link has been considered. This will be the subject of (near) future publications.

A bit more specifically, in order to give a quantitative description of information acquisition in a general setting, the process of information acquisition needs to be formalized. We do it in a reasonably obvious way: by introducing notions of questions that an agent (who is assumed to be solving a certain problem) can ask an information source and answers that the information source can provide in response.
The central concept introduced in \cite{part1} is that of {\it question difficulty} which is a real-valued functional defined on the set of questions. The meaning of question difficulty is in that a given source can provide more accurate answers to questions with lower difficulty. This article develops a symmetric concept of {\it answer depth} which is a functional on the set of answers that can be informally thought of as a measure of the amount of ``work'' a source has to do to provide the answer. If the depth is close to the corresponding question's difficulty, the answer is very accurate and vice versa. Thus, for a given question, more accurate answers have larger values of depth. On the other hand, sources can be characterized with {\it capacity} \footnote{Note that this is not {\it information} capacity like that of channel in classical Information Theory, but rather {\it pseudoenergy} capacity.} that describes the highest answer depth a source can provide in response to any question.

Both question difficulty and answer depth are source-specific objects that describe the source {\it knowledge structure}. Assuming that the description is faithful, a source should be expected to give answers of equal depth to questions of equal difficulty, regardless of the details of these questions \footnote{Finding this to be false for some questions would imply that the current description of the source's knowledge structure is not sufficiently accurate and that, for example, a more elaborate model is needed.}. This implies that, for a given source, the answer depth should be a function of question difficulty. We refer to such a function as a {\it source model}. The specific shape of the source model would in general have to be found experimentally. However, some features of realistic source models can be predicted from principles of ``consistency with general experience''. We discuss such considerations in Section~\ref{s:sources}.

\subsection{Related work}
 This article, together with \cite{part1}, is an integral part of an effort to extend Information Theory from the realm of communications into that of information acquisition and usage for solving problems. So it can be looked upon as an extension of classical Information Theory which, as was mentioned earlier, is focused primarily on information quantity, with Shannon entropy being the central concept involved in proper quantification of the latter.
 Besides fundamental advances in communications, the list of successful application of this and derivative concepts includes
(but by no means is limited to) new algorithms in computer vision \cite{viola1995}, new methods of analysis in climatology \cite{mokhov2006,verdes2005},
physiology \cite{katura2006} and neurophysiology \cite{chavez2003}. The concept of pseudoenergy introduced in \cite{part1} and used in the present article extends that of entropy in the direction of information content and provides the foundation for the quantitative description of knowledge possessed by various information sources.

The idea of using additional information to improve the decision quality has been studied in the area of statistical decision making. One can mention applications to innovation
adoption \citep{mccardle1985,jensen1988}, fashion decisions \citep{fisher1996} and vaccine composition decisions for flu immunization \citep{kornish2008}. Typically, the amount of information in these applications is measured simply by the number of relevant observations of certain random variable realizations. Some authors \cite{fischer1996,ellison1993} introduced models
(for instance, the effective information model) for accounting for the actual amount of
information  contained in the received observations.

 The problem of optimal usage of information obtained from experts has been addressed mostly in the form of updating the decision maker's beliefs given probability assessment from multiple experts \cite{french1985,genest1986,clemen1987,clemen1999} and, in particular, optimal combining of expert opinions, including experts with incoherent and missing outputs \cite{predd2008}. In particular, investigations on combining information of experts that partition the event differently \cite{bordley2009} and on rules of updating probabilities based on outcomes of partially similar events \cite{bordley2011} are close in spirit to the approach developed here in that they deal with different types of information. The emphasis of the proposed approach is on {\it optimizing} on the particular type of information and on the explicit consideration of the dependence of the optimal information on both for the expert and the decision making problem.

 This article uses an axiomatic approach to determine the overall form of the answer depth functional. The latter, together with the related concept of question difficulty studied in \cite{part1}, can be thought of as a logical development of the entropy concept of information theory. The axiomatic approach was  used in \cite{faddeev1956} to derive the most general form of the (Shannon) entropy function. A different set of axioms was used in  \cite{RENYI:1961} to find the one-parameter family of functions (known as R\'enyi entropies) that included standard entropy as a special case.  The concept of structural entropy was introduced in \cite{havrda1967} and used for classification purposes. The Havrda-Charvat entropy was derived by axiomatic means in \cite{simovici2002} where axiomatization of partition entropy was discussed on rather general grounds (see also \cite{simovici1999}).

Information Physics (see \cite{caticha-rev} for a recent fairly comprehensive review) is a relatively recently developed branch of physical sciences focused on the role of information in fundamental laws of physics. It is fair to say that Information Physics dates back to the original work of Jaynes \cite{JAYNES:1957a,JAYNES:1957b} on classical and quantum thermodynamics. There it was shown that the main laws of the latter could be derived from maximization of Shannon entropy subject to appropriate constraints expressing the macrostate parameters. These results were later extended to derivations of classical \cite{caticha07} and quantum \cite{caticha11} mechanics main laws. Recently, progress also has been made in obtaining main equations of relativistic quantum theory \cite{caticha12}. The central hypothesis of Information Physics is that the fundamental physics laws are indeed the laws of inductive inference applied to the description of respective systems. The main emphasis in discovering fundamental laws thereby shifts to that of determining the correct degrees of freedom and the relevant information necessary for the description of the system state. From the point of view of main information attributes, it can be said that, while the classical Information Theory's main concern is with information quantity, Information Physics' focus is on information relevance.

A somewhat different direction within the field of Information Physics exploits the consequences of theory of partially ordered sets (posets). This direction goes back to the work of Cox \cite{cox1946,cox1961,cox1979} on foundations of probability considered as a way to consistently describe incomplete information a conscious agent may possess. More recently, it has been shown in \cite{knuth05,knuth07,knuth08} that while probability is a natural (bi-)valuation on the lattice of logical assertions about system states, Shannon entropy (which is the main tool of information quantification in classical Information Theory) is a natural (bi-)valuation on the corresponding lattice of questions. It has been argued that order may be one of the most fundamental concepts of science and it has been demonstrated \cite{knuth-sr} that, for instance, Lorentz transformations and Minkowski metric of special relativity can be derived directly from order-theoretic considerations applied to events in space-time.

\subsection{Outline}
The rest of the article is organized as follows. In Section~\ref{s:prelim}, we briefly discuss the necessary preliminaries. In Section~\ref{s:frame}, the main primitives of the proposed framework -- questions and answers -- are defined. In Section~\ref{s:depth}, the overall form of the answer depth functional is derived from a set of plausible postulates that express, in particular, the isotropy property of the source's knowledge structure. Section~\ref{s:diff-depth} describes relationships between question difficulty and answer depth for main types of possible questions. In Section~\ref{s:quasi-perfect}, a special class of answers -- the quasi-perfect answers -- is discussed. Section~\ref{s:relations} is devoted to relationships between different questions and, in particular, the relative depth of an answer to one question with respect to another question is introduced.
Section~\ref{s:sources} introduces the notion of a source model and proposes several simple models characterized by a well-defined source capacity.  Section~\ref{s:estimation} discusses optimization-based methods for estimating both the source knowledge structure and source model parameters.
Section~\ref{s:examples} gives simple numerical examples illustrating concepts and results discussed earlier in the article. Finally, Section~\ref{s:conclusion} gives a short summary and a discussion of main results.


\section{\label{s:prelim}Preliminaries}
The necessary preliminary facts and definitions were already discussed in the companion paper \cite{part1}.
We briefly recap it here for so that the present article can be read independently. If uncertainty is present in a decision making problem, it can de described as a certain base space $\Omega$ (equipped with a suitable sigma-algebra $\mF$) that contains all possible sets of input data for the problem. The problem itself can be formulated as an optimization with respect to a suitably chosen criterion.

When uncertainty is present, a notion of {\it loss} can usually be defined. It measures the performance of a solution obtained in the presence of uncertainty with respect to that of a solution that would have been obtained had the decision maker possessed the full information. The overall goal of the agent can be formulated as that of minimizing the expected loss. To achieve that goal, the agent can turn to an information source for additional information.

We call a collection of (distinct) subsets $\bC=\{C_1,\dotsc, C_r\}$ {\it inclusion-free} if for any $C_i, C_j\in \bC$ neither of the two is a proper subset of the other. A collection of subsets  $\bC=\{C_1,\dotsc, C_r\}$ is called {\it complete} if $\cup_{i=1}^r  C_i =\Omega$.

A {\it complete partition} $\bC=\{C_1,\dotsc, C_r\}$ of $\Omega$ is a collection of (measurable) subsets $C_j\in \mF$ of $\Omega$ such that $C_j\cap C_l=\emptyset$ for $j\ne l$ and $\cup_{j=1}^r C_j=\Omega$. A partition $\tilde\bC$ is a {\it refinement} of $\bC$ if every set from $\tilde\bC$ is a subset of some set from $\bC$. In such a case, $\bC$ is a {\it coarsening} of $\tilde\bC$.

\begin{figure}[hbt]
\includegraphics[scale=0.6]{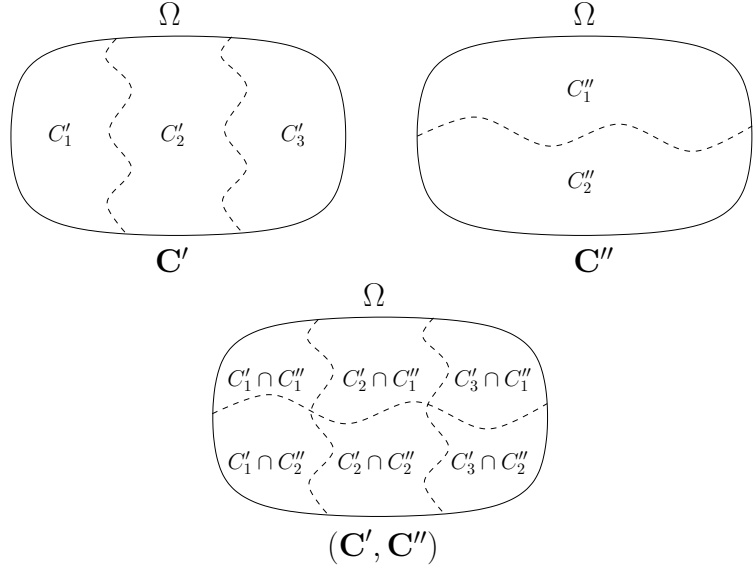}
 \caption{\label{f:partitions}Two (complete) partitions of $\Omega$ and the corresponding joint partition.}
\end{figure}

 If $\bC'=\{C_1',\dotsc, C_r'\}$ and $\bC''=\{C_1'',\dotsc, C_s''\}$ are two partitions of $\Omega$ then the partition $\bC=\bC'\cap \bC''$ is defined as the partition that consists of all sets of the form $C_i'\cap C_j''$: $\bC'\cap \bC''=\{C_1'\cap C_1'', C_1'\cap C_2'', \dotsc, C_r'\cap C_s''\}$ (see Fig.~\ref{f:partitions} for an illustration).  Clearly,  $\bC'\cap \bC''$ is a refinement of both $\bC'$ and $\bC''$.

\begin{figure}[hbt]
\includegraphics[scale=0.6]{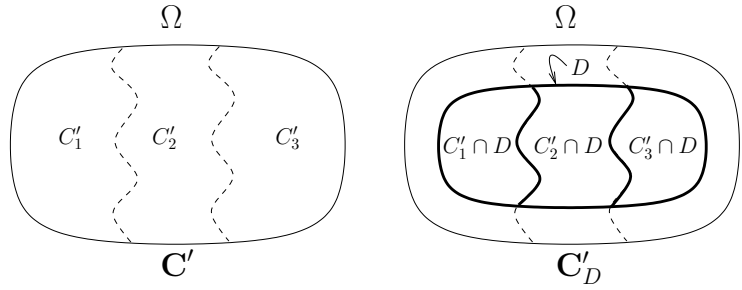}
\caption{\label{f:partition_induced}Partition $\bC_D$ of set $D\subset\Omega$ induced by a partition $\bC$ of $\Omega$.}
\end{figure}

If $D\subset \Omega$ is a subset of $\Omega$ and $\bC'=\{C_1',\dotsc, C_r'\}$ is a partition of $\Omega$, the partition $\bC'_D=\{D\cap C_1',\dotsc, D\cap C_r'\}$ of $D$ will be called the partition of $D$ {\it induced} by the the partition $\bC'$ of $\Omega$ (see Fig.~\ref{f:partition_induced}).

Besides complete partitions of $\Omega$, we also make use of {\it incomplete partitions} $\bC=\{C_1,\dotsc, C_r\}$ such that $\cup_{i=1}^r C_i\ne \Omega$. For any partition $\bC$, we use the notation  $\hat C\equiv \cup_{i=1}^r C_i$. Clearly, partition $\bC$ is complete if and only if $\hat C = \Omega$.

For an arbitrary complete partition $\bC=\{C_1,\dotsc, C_r\}$, the measure $P$ can be written as a linear combination of conditional measures as
\begin{equation}
P=\sum_{j=1}^r P(C_j)P_{C_j},
\label{eq:P-decomp}
\end{equation}
where a conditional measure $P_C$ is defined, for any subset $C$ of $\Omega$ such that $P(C)>0$, by
\begin{equation}
P_C(D)=\frac{P(D\cap C)}{P(C)},
\label{eq:cond-measure}
\end{equation}
for arbitrary $D\in \mF$.

\section{\label{s:frame}Information acquisition primitives: questions and answers}
As was stated earlier, the basic information acquisition process involves the agent asking questions of an information source \footnote{It should be noted at this point that the role of information source can be played by conscious agents (human experts) and various data sources alike. In the latter case, additional care has to be taken interpreting questions and answers but the overall construction still applies. This theme will be developed further in future publications.} and the source providing answers. Here we recall the definition of questions discussed in \cite{part1} and provide a definition of answers to be used later in this article.

\subsection{Questions}
A question was originally defined in \cite{cox1979} as a set of logical assertions that answer it. A somewhat different notion of a question was proposed in \cite{caticha04} where questions were identified with probability distributions that are interpreted as requests for missing information.
The direction suggested in \cite{cox1979} was further pursued in \cite{knuth05,knuth08} where a distributive lattice of questions defined as down-sets of (sets of) logical assertions was described.
Our definition of questions described in detail in the companion paper \cite{part1} combines features of those proposed in \cite{cox1979,knuth05} on one hand and in \cite{caticha04}, on the other.
  Specifically, questions are associated with inclusion-free collections of subsets of the base set $\Omega$ which, as mentioned in the previous section, comes ``equipped'' with a probability measure $P$ describing the ``initial state of information'' available to the agent. Depending on the nature of the particular collection of subsets, several different types of questions can result. Thus, for a single subset, one obtains an {\it ideal question} of \cite{knuth05}, a complete partition corresponds to a {\it partition question}, and a finest partition (in case it exits) yields the {\it central issue}, i.e. the most detailed partition question. Furthermore, a complete collection of subsets of $\Omega$ (regardless of it being a partition) corresponds to a {\it real question} in terminology of \cite{cox1979} and \cite{knuth05}. These are the questions that lie above the central issue in the distributive lattice of questions. The questions of interest to us at this point \footnote{It is possible that other questions will play a role in future developments.} will be those corresponding to partitions of $\Omega$ -- both complete and incomplete. We will refer to them as {\it complete} and {\it incomplete} questions, respectively.
Thus a question in what follows is identified with a partition $\bC=\{C_1, C_2,\dotsc, C_r\}$ of $\Omega$. The questions for which $\hat C=\Omega$ are called {\it complete} questions. In particular, incomplete questions for which the corresponding partition consists of a single set $C\subset \Omega$ will often be  called, following \cite{knuth05}, {\it ideal} questions. We will also use the terms ``question'' and ``partition'' interchangeably.

A {\it difficulty functional} $G(\Omega,\bC,P)$ can be associated with any question $\bC$. The particular form of $G(\Omega,\bC,P)$ can be determined if some requirements, which can be formulated as {\it postulates}, are imposed. This was done in the companion paper \cite{part1} where a particular system of postulates that embodied {\it linearity} and {\it isotropy} properties of the source's {\it knowledge structure} and, hence, the difficulty functional, was proposed. The main theorem proved in \cite{part1} derives the general form of the difficulty functional that is required to satisfy such postulates.

\begin{theorem}
Let the functional $G(\Omega, \bC, P)$ where $\bC=\{C_1,\dotsc, C_r\}$ satisfy Postulates 1 through 6 (see \cite{part1}). Then it has the form
$$G(\Omega, \bC, P)=\frac{\sum_{j=1}^r u(C_j)P(C_j)\log \frac{1}{P(C_j)}}{\sum_{j=1}^r P(C_j)},$$
where $u(C_j)=\frac{\int_{C_j}u(\o)\,dP(\o)}{P(C_j)}$ and $u$: $\Omega\rightarrow \bR$  is an integrable nonnegative function on the parameter space $\Omega$.
\label{th:G}
\end{theorem}

In particular, for the given question $\bC$, its difficulty depends, besides the initial probability measure $P$, on the (integrable) function $u(\cdot)$ defined on the parameter space $\Omega$. This function was called the {\it pseudotemperature} in \cite{part1} using parallels with thermodynamics (see \cite{part1} for more details). The question difficulty then can be interpreted as the amount of {\it pseudoenergy} associated with question $\bC$.

\subsection{Answers}
Given a question $\bC$, a source is assumed to be capable of providing an answer. Our definition of an answer differs somewhat from that proposed in \cite{cox1979} in that it aims to account for different degrees of accuracy. Also, as opposed to the definition of Cox from \cite{cox1979}, according to which an answer is a logical assertion that answers {\it no less than} the question asked, we take special care to make sure that {\it no more than} the question is being answered.

Since any information is represented by some measure on $\Omega$, it is reasonable to think of an answer to question $\bC$ as a message a reception of which implies certain changes in the initial measure $P$. In an extreme case, a message can change the original measure to a measure supported at a single element of $\Omega$ -- this describes a complete resolution of the initial uncertainty and the best possible answer to the exhaustive question which is the central issue of \cite{knuth05}.
Taking these considerations into account, and assuming, without loss of generality, that $P(C_j)>0$ for all subsets $C_j$ in $\bC$, we adopt the following definition.

{\bf Definition:} An answer to the question $\bC=\{C_1,\dotsc, C_r\}$ is a message $V(\bC)$ that takes values in the set $\{s_1,s_2,\dotsc, s_m\}$ such that a reception of the value $s_k$ of the message updates the initial measure $P$ on $\Omega$ to the measure $P^k$ such that either $P^k(C_j)=0$ or $P^k_{C_j}=P_{C_j}$ for all $k=1,\dotsc m$ and all $j=1,\dotsc, r$.

The meaning of the condition $P^k_{C_j}=P_{C_j}$ in this definition is that an answer to question $\bC$ should not resolve more uncertainty than what the question $\bC$ was requesting: a valid answer $V(\bC)$ does not change the relative probabilities {\it inside} subsets $C_j$, but only the probabilities {\it between} the subsets $C_j$ constituting the question.
It is straightforward to show for $V(\bC)$ to be an answer to a complete question $\bC$ according to this definition, it is necessary and sufficient for the updated measures $P^k$, $k=1,\dotsc, m$, to take the form
\begin{equation}
P^k=\sum_{j=1}^r p_{kj}P_{C_j},
\label{eq:Pk-mc}
\end{equation}
where $p_{kj}$, $k=1,\dotsc, m$, $j=1,\dotsc, r$ are nonnegative coefficients such that $\sum_{j=1}^r p_{kj}=1$ for $k=1,\dotsc, m$.

If question $\bC$ is complete and $V(\bC)$ is a corresponding answer, we assume that $V(\bC)$ does not change the original measure $P$ on average, or, formally speaking,
\begin{equation}
\sum_{k=1}^m \Pr(V(\bC)=s_k) P^k = P,
\label{eq:cons-mc}
\end{equation}
from which it follows, in particular, that if the answer is perfect, then $\Pr(V^*(\bC)=s_j)=P(C_j)$. We refer to (\ref{eq:cons-mc}) as the {\it consistency with prior} condition for the answer $V(\bC)$.

In the following, we denote the probability $\Pr(V(\bC)=s_k)$, for any answer $V(\bC)$ to a complete question $\bC$ -- by $v_k$, for brevity. It is straightforward to show that it follows from the consistency with prior condition (\ref{eq:cons-mc}) that
\begin{equation*}
 \sum_{k=1}^m v_kp_{kj} = P(C_j),\;\; j=1,\dotsc, r.
\end{equation*}

Incomplete questions, including ideal questions, are interpreted, as explained in \cite{part1}, as
``aspects'' of complete questions conditioned on the corresponding subsets of $\Omega$ being true statements. For example, if the complete question is {\it ``Is the fruit an apple, a pear or a peach?''} then the ideal question corresponding to the element (subset) {\it ``Apple''} of the base space can be understood as the complete question provided the fruit is really an apple. Thus ideal and, more generally, incomplete, questions are never formulated and posed as such \footnote{For instance, in \cite{caticha04}, an example of an ideal question is given: {\it ``What color is Napoleon's white horse?''}. In our interpretation, an ideal question corresponding to the element {\it ``White''} of the base space $\Omega$ consisting of all possible horse colors would not be phrased this way. Rather, the verbalized question will simply be {\it ``What color is Napoleon's horse?''} and the particular ideal question will be implicitly asked if and only if Napoleon's horse is indeed white. Obviously, the latter fact (that the ideal question being asked corresponds to the element {\it ``White''}) can only be known to someone having certain knowledge of the true color of Napoleon's horse.} in real inquiry situations. On the other hand, when a real question is posed, it is always true that an ideal question is implicitly asked. Often, though, neither the agent nor the source (unless the source is capable of providing perfect answers) actually know which ideal question is being asked. The reason is that, simply, the knowledge of the correct answer \footnote{Here, just like in \cite{part1}, we use the term ``correct answer'' for a complete question $\bC$ to denote the subset $C\in \bC$ such that $\o\in C$ for the given instance of the question.  Using this terminology, a perfect answer to a complete question is simply a message that identifies the correct answer with certainty. Any ideal question has a single correct answer and is fully described by the latter.} to the given question is needed for that.

Let $\bC=\{C_1,\dotsc, C_r\}$ be a complete question and let $C_j\in \Omega$ be one of its ``constituent'' ideal questions. Suppose $V(\bC)$ is some answer to $\bC$. We denote by $q_k^{(j)}$ the probability that the corresponding answer to $C_j$ takes the value $s_k$. Then one can apply the Bayes' rule to obtain that
\begin{equation}
q_k^{(j)}=\Pr(V(\bC)=s_k|\o\in C_j)=\frac{p_{kj}v_k}{P(C_j)}.
\label{eq:qkj}
\end{equation}
In particular, if the answer $V(\bC)$ is perfect, then it follows from (\ref{eq:qkj}) (since $p_{kj}=\delta_{kj}$ and $v_j=P(C_j)$) that $q_k^{(j)}=\delta_{kj}$, i.e. every ideal question $C_j$ receives just a single answer -- equal to its correct answer.

To consider a more general incomplete question, assume, without loss of generality, that such a question has the form $\bC'=\{C_1,\dotsc, C_l\}$ where $l<r$. If, just like above, $V(\bC)$ is some answer to the complete question $\bC=\{C_1,\dotsc, C_r\}$, the probabilities $\hat q_k$ of different values of the ``induced'' answer to $\bC'$ can be found by conditioning on $\hat C=\cup_{j=1}^l C_j$:
\begin{equation}
\hat q_k=\Pr(V(\bC)=s_k|\o\in \hat C)=v_k\frac{P^k(\hat C)}{P(\hat C)}.
\label{eq:hat-qk}
\end{equation}

In the following, we will sometimes refer to incomplete (including ideal) questions and answers to them without a simultaneous explicit reference to the corresponding complete (and thus real) question. It has to be kept in mind, however, that incomplete questions and corresponding answers are auxiliary constructs in the sense described above.

While the functional $G(\Omega, \bC, P)$ measures difficulty of questions,
it would be desirable to develop a measure of the amount of difficulty in $\bC$ that is resolved by the answer $V(\bC)$. As was mentioned earlier, the question difficulty can be interpreted as the amount of {\it pseudoenergy} associated with the question. It is reasonable to expect the amount of pseudoenergy contained in a perfect answer to be equal to that in the question itself and, respectively, the amount of pseudoenergy in any other answer to contain somewhat less pseudoenergy -- as long as it is an answer to $\bC$ and not some other question.

In the following we denote the amount of pseudoenergy contained in the answer $V(\bC)$ -- the {\it depth} of $V(\bC)$ -- by $Y(\Omega, \bC, P, V(\bC))$ to emphasize its dependence on $\Omega$ and the initial measure $P$.

\section{\label{s:depth}Answer depth functional}
In this section, our goal is to derive the general form of the answer depth functional by imposing certain plausible requirements it has to satisfy. These requirements that we call Postulates are similar
to those stated in Postulates Q1 through Q6 for questions (see \cite{part1}).

Since information in $V(\bC)$ is conveyed by means of modifying the original measure $P$ and the latter is modified differently for each value of the message $V(\bC)$  the depth functional for the answer $V(\bC)$ should be equal to the expected value over possible values of the message  $V(\bC)$:
\begin{equation}
Y(\Omega, \bC, P, V(\bC)) = \sum_{k=1}^m \Pr(V(\bC)=k) Y(\Omega, \bC, P, P^k),
\label{eq:Ydef}
\end{equation}
where $P^k$ is the measure modified by the reception of $V(\bC)=k$ and $Y(\Omega, \bC, P, P^k)$ is the conditional depth that depends on the modified measure $P^k$.

We now impose some reasonable requirements on conditional depth functionals $Y(\Omega, \bC, P, P^k)$ which are formulated as postulates as in \cite{part1}.

The first such requirement is that the conditional depth should vanish if the measure is not modified at all, i.e. if $P^k=P$. On the other hand, if the modified measure assigns larger probabilities to {\it all} subsets in $\bC$ (which can happen only for incomplete questions), then the conditional depth should be strictly positive. This is the content of Postulate A1.

{\bf Postulate A1} ({\it Correct direction}). Let $\bC=\{C_1,\dotsc, C_r\}$ be any question. Then $Y(\Omega, \bC, P, P^k)= 0$ if $P^k(C_j)=P(C_j)$ for all $j=1,\dotsc r$ and $Y(\Omega, \bC, P, P^k)> 0$ if $P^k(C_j)>P(C_j)$ for all $j=1,\dotsc r$.

The second part of the postulate says that, for an ideal question, if, for instance, upon reception of the value $s_k$ of $V(\bC)$ the set $C$ has a higher probability than before then the value $k$ has a positive amount of pseudoenergy. For example, if the original question was {\it ``What kind of fruit is it?''} with {\it ``Pear''} being the correct answer then in case the answer sounds like {\it ``It looks a lot like a pear''} or {\it ``It's either a pear or an apple''}  then such answer is assigned positive pseudoenergy as it moves ``in the right direction'' towards the correct answer.

The next postulate parallels Postulate Q2 for questions (see \cite{part1}).

{\bf Postulate A2} ({\it Continuity}). The function $Y(\Omega, \bC, P, P^k)$ is continuous in all parameters it may depend upon.

The next postulate follows from the requirement that if $V(\bC)$ is an answer to question $\bC$ then the depth of $V(\bC)$ cannot exceed the difficulty of $\bC$. This property is easiest to state for ideal questions $C\subset \Omega$.

{\bf Postulate A3} ({\it Ideal complete answer}). Let $C$ be an ideal question and suppose $P^k(C)=1$. Then $$Y(\Omega, C, P, P^k) = G(\Omega, C, P). $$

This postulate expresses a simple desideratum that an exhaustive correct answer to a question should convey exactly the amount of information requested by the question. For instance, if the question is {\it ``What fruit is it?''} with {\it ``Apple''} as a correct answer then the answer {\it ``Apple''} should carry all the information the question was asking for.

The next three postulates parallel Postulates Q3 through Q5 for questions.

{\bf Postulate A4} ({\it Incomplete question answer decomposition}) Let $\bC=\{C_1,\dotsc, C_r\}$ be an incomplete question.
 Then $$Y(\Omega, \bC, P,P^k)=Y(\Omega, \hat C, P,P^k)+Y(\hat C, \bC, P_{\hat C}, P^k_{\hat C}). $$

{\bf Postulate A5} ({\it Mean value}). Let $\bC$ and $\bC'$ be two incomplete questions such that $\hat C \cap \hat C'=\emptyset$. Then
 $$Y(\Omega, \bC\cup \bC', P, P^k)= \frac{P^k(\hat C)Y(\Omega, \bC, P, P^k)+P^k(\hat C')Y(\Omega, \bC', P,P^k)}{P^k(\hat C \cup \hat C')}. $$

 Just like we did for questions, we can say that the subset $D$ of $\Omega$ is  {\it homogeneous} iff the conditional depth functional depends only on measures of partition $\bC$ whenever $\hat C\subset D$, i.e. $Y(D, \bC, P_D, P^k_D)= f(P_D(\bC), P_D^k(\bC))$. In particular, any atom (minimal set) of $\mF$ is homogeneous.

{\bf Postulate A6} ({\it Homogeneous ideal sequentiality}). Let $D\subseteq \Omega$ be a homogeneous subset of the parameter space and let $\bC$ be a question such that $\hat C\subset D$. Then
 $$Y(\Omega, C, P,P^k)=Y(\Omega, D, P,P^k) + Y(D,C,P_D,P^k_D). $$

 We can now state the main result about the possible shape of answer conditional depth functional $Y(\Omega, \bC, P,P^k)$. It is formulated as a theorem.

 \begin{theorem}
 Let Postulates A1 through A6 hold. Then the conditional answer depth functional $Y(\Omega, \bC, P,P^k)$ has the following form
 $$Y(\Omega, \bC, P,P^k)=\frac{\sum_{j=1}^r u(C_j)P^k(C_j)\log \frac{P^k(C_j)}{P(C_j)}}{\sum_{j=1}^r P^k(C_j)}, $$
 where $u(C_j)=\frac{\int_{C_j} u(\o)dP^k(\o)}{P^k(C_j)}$ and the integrable function $u$: $\Omega\rightarrow \bR$ is the same that is used in characterizing the question difficulty functional $G(\cdot)$.
 \label{th:Ycond}
 \end{theorem}

 {\bf Proof:} The proof is similar to that of main theorem in \cite{part1}.
 Let $\bA=\{A_1,\dotsc, A_M\}$ be a complete and sufficiently fine partition of $\Omega$. We can assume, without loss of generality, that the sigma-algebra $\mF$ on $\Omega$ is comprised of all unions of sets in $\bA$. Note, in particular, that all subsets in $\bA$ are homogeneous.

 Let $D$ be a homogeneous subset of $\Omega$ and let $C'\subset C\subset D$ be two subsets of $D$. Then, by Postulate A6,
 \begin{equation}
 Y(\Omega, C, P, P^k)=Y(\Omega, D, P, P^k)+Y(D, C, P_D, P^k_D),
 \label{eq:YOmDC}
 \end{equation}
 and
 \begin{equation}
 Y(D, C', P_D, P^k_D)=Y(D, C, P_D, P^k_D)+Y(C, C', P_C, P^k_C).
 \label{eq:YDCC'}
 \end{equation}
 Since $D$ is homogeneous it follows from (\ref{eq:YDCC'}) that
 \begin{equation*}
 f(P_D(C'),P^k_D(C'))=f(P_D(C),P^k_D(C))+f(P_D(C')/P_D(C), P^k_D(C')/P^k_D(C)).
 \label{eq:f(p,q)}
 \end{equation*}
 Then standard arguments using Postulates A1 and A2 (see \cite{RENYI:1961} for details) lead to the conclusion
 that the function $f(\cdot)$ has the form
 \begin{equation*}
 f(p,q)=c\log \frac{q}{p},
 \end{equation*}
 where $c$ is a positive constant. Going back to $Y$ we obtain
 \begin{equation}
 Y(D, C, P_D, P^k_D)= u'(D) \log \frac{P^k_D(C)}{P_D(C)},
 \label{eq:YDCres}
 \end{equation}
 where $u'(D)>0$ is a constant that can possibly depend on the particular homogeneous subset $D$.

 Substituting (\ref{eq:YDCres}) into (\ref{eq:YOmDC}) we arrive at
 \begin{align*}
 Y(\Omega, C, P, P^k)-Y(\Omega, D, P, P^k)&= Y(D, C, P_D, P^k_D)=u'(D)\log \frac{P^k_D(C)}{P_D(C)}\\
 &= u'(D)\log \frac{P^k(C)}{P(C)}-u'(D)\log \frac{P^k(D)}{P(D)},
 \end{align*}
 from which it follows (using continuity of $Y$ and the fact that the subset $C\subset D$ is arbitrary) that
 \begin{equation*}
 Y(\Omega, C, P, P^k)= u'(D) \log \frac{P^k(C)}{P(C)} + v'(D),
 \end{equation*}
 for any $C\subset D$ whenever $D$ is homogeneous. Here $v'(D)$ is another constant that can possibly depend on the homogeneous subset $D$. We can now use Postulate A3 to conclude that $u'(D)=u(D)$ for all homogeneous sets $D$ and that $v(D')\equiv 0$. This leads to the following expression for the conditional depth functional of an answer to an ideal question lying inside a homogenous subset:
 \begin{equation}
 Y(\Omega, C, P, P^k)= u(D) \log \frac{P^k(C)}{P(C)}.
 \label{eq:YOmCres}
 \end{equation}

 Now let $\bD=\{D_1,\dotsc, D_N\}$ be a (complete) partition of $\Omega$ such that every subset in $\bD$ is homogeneous. Let $C\subset \Omega$ be an ideal question. Using Postulate A5, we can write
 \begin{equation}
 Y(\Omega, \bD_C, P, P^k)=\frac{\sum_{j=1}^N u(D_j)P^k(C\cap D_j)\log \frac{P^k(C\cap D_j)}{P(C\cap D_j)}}{P^k(C)},
 \label{eq:YOmD_CP}
 \end{equation}
 and
 \begin{equation}
 Y(C, \bD_C, P_C, P^k_C)=\sum_{j=1}^N u(D_j)\frac{P^k(C\cap D_j)}{P^k(C)}\log \frac{P^k(C\cap D_j)/P^k(C)}{P(C\cap D_j)/P(C)}.
 \label{eq:YCD_CP_C}
 \end{equation}
 An application of Postulate A4 now yields
 \begin{align*}
 Y(\Omega, C, P, P^k)&= Y(\Omega, \bD_C, P, P^k)-Y(C, \bD_C, P_C, P^k_C)\\
 &= \sum_{j=1}^N u(D_j)\frac{P^k(C\cap D_j)}{P^k(C)}\log \frac{P^k(C)}{P(C)}= u(C)\log \frac{P^k(C)}{P(C)},
 \end{align*}
 where
 \begin{equation}
 u(C)\equiv \sum_{j=1}^N \frac{P^k(C\cap D_j)u(D_j)}{P^k(C)}=\frac{\int_C u(\o) dP^k(\o)}{P^k(C)}.
 \label{eq:U(C)ans}
 \end{equation}
 Here, the function $u$: $\Omega\rightarrow \bR$ is defined
 as
 $$u(\o)=\sum_{j=1}^N u(D_j) I_{D_j}(\o),$$
 and therefore is the same exact function that was used to describe the question difficulty functional~$G$.

 Finally, let $\bC=\{C_1,\dotsc, C_r\}$ be an arbitrary question on $\Omega$. An application of Postulate~A5 yields
 \begin{equation}
 Y(\Omega, \bC, P, P^k)=\frac{\sum_{j=1}^r u(C_j)P^k(C_j)\log \frac{P^k(C_j)}{P(C_j)}}{\sum_{j=1}^r P^k(C_j)},
 \end{equation}
 where $u(C_j)$ is given by (\ref{eq:U(C)ans}). \qed

 Having found the expression for conditional depth functional we can now use it to obtain the unconditional (expected) answer depth $Y(\Omega,\bC,P,V(\bC))$. We formulate the result as a corollary.
 \begin{corollary}
 The answer depth functional $Y(\Omega, \bC, P, V(\bC))$ has the form
 $$Y(\Omega, \bC, P, V(\bC))=\sum_{k=1}^m \Pr(V(\bC)=s_k)\frac{\sum_{j=1}^r u(C_j)P^k(C_j)\log \frac{P^k(C_j)}{P(C_j)}}{\sum_{j=1}^r P^k(C_j)}, $$
 where $P^k$ is the measure on $\Omega$ updates by a reception of $V(\bC)=s_k$ and
 $u(C_j)$ is defined in Theorem~\ref{th:Ycond}.
 \label{c:Y}
 \end{corollary}

 \section{\label{s:diff-depth}Relationships between difficulty and depth}
 Theorems~\ref{th:G} and \ref{th:Ycond} (together with Corollary~\ref{c:Y}) establish the overall form that question difficulty and answer depth, respectively, can take. The conditional depth functional $Y(\Omega, \bC, P, P^k)$ depends, besides the original measure $P$, on the updated measure $P^k$.

 \subsection{Complete questions}
 Let us assume the consistency with prior condition (\ref{eq:cons-mc}) holds and consider the answer depth functional given by Corollary~\ref{c:Y}. Since for a complete question $\sum_{j=1}^r P^k(C_j)=1$, we can write
 \begin{align*}
 & Y(\Omega,\bC,P, V(\bC)) = \sum_{k=1}^m v_k\sum_{j=1}^r u(C_j)P^k(C_j)\log \frac{P^k(C_j)}{P(C_j)} \\
 &= \sum_{k=1}^m v_k\sum_{j=1}^r u(C_j)P^k(C_j)\log P^k(C_j) -
 \sum_{k=1}^m v_k\sum_{j=1}^r u(C_j)P^k(C_j)\log P(C_j) \\
 &\overset{(a)}{=} \sum_{k=1}^m v_k\sum_{j=1}^r u(C_j)P^k(C_j)\log P^k(C_j) +
 G(\Omega,\bC,P)\overset{(b)}{\le } G(\Omega,\bC,P),
 \end{align*}
 where (a) follows from (\ref{eq:cons-mc}) and Theorem~\ref{th:G}, and (b) follows from the inequality $\log P^k(C_j)\le 0$. It is also clear that the inequality (b) becomes an equality if and only if, for every value $s_k$ of the answer message, either $P^k(C_j)=0$ or $\log P^k(C_j)=0$ for every value of the index $j$. For the latter to be true it is necessary and sufficient that, for all values of $k$,
 \begin{equation}
 P^k(C_j)=\delta_{f(k),j},
 \label{eq:Pk(Cj)}
 \end{equation}
 where $f$: $\{1,2,\dotsc, m\}\rightarrow \{1,2,\dotsc, r\}$ is a map from the set of possible values of index $k$ to that of index $j$. Substituting (\ref{eq:Pk(Cj)}) into (\ref{eq:cons-mc}) we obtain
 \begin{equation}
 P(C_j)=\sum_{k=1}^m v_k \delta_{f(k),j}=\sum_{k:f(k)=j}v_k.
 \label{eq:P(Cj)}
 \end{equation}
 It is easy to see that without loss of generality one can define an equivalent message $V'(\bC)$ such that $V'(\bC)=s_j$ whenever $V(\bC)=s_k$ such that $f(k)=j$. Then (\ref{eq:P(Cj)}) becomes simply
 \begin{equation}
 P(C_j)=\Pr(V'(\bC)=s_j).
 \label{eq:P(Cj)perf}
 \end{equation}

Recall that a perfect answer to a complete question $\bC=\{C_1,\dotsc, C_r\}$ is defined as a message $V^*(\bC)=\{s_1,\dotsc, s_r\}$ such that $P^k(C_j)=\delta_{kj}$, and, as a consequence, $\Pr(V(\bC)=s_j)=P(C_j)$. Then we can state the result obtained above as a lemma.

 \begin{lemma}
 Let $\bC$ be a complete question and assume the condition (\ref{eq:cons-mc}) for any answer $V(\bC)$ to $\bC$ to hold. Then $Y(\Omega,\bC,P,V(\bC))\le G(\Omega,\bC,P)$ with the inequality being tight if and only if the answer $V(\bC)$ is perfect (up to trivial equivalences).
 \label{l:ans-mc}
 \end{lemma}

 \subsection{Ideal and other incomplete questions}
 Let $C\subset \Omega$ be an ideal question. We can write the depth functional for a corresponding answer (denoting by $q_k$ the probability that $V(C)=s_k$) $V(C)$ as follows.
 \begin{align*}
 Y(\Omega,C,P,V(C))&= \sum_{k=1}^m q_k u(C)P^k(C)\log \frac{P^k(C)}{P(C)} \\
                   &= u(C) \sum_{k=1}^m  q_k\log P^k(C) - u(C)\log P(C) \sum_{k=1}^m \Pr(V(C)=s_k) \\
                   &= u(C) \sum_{k=1}^m  q_k\log P^k(C) + G(\Omega, C, P) \overset{(a)}{\le} G(\Omega, C, P),
 \end{align*}
 where (a) follows from that the inequality $\log P^k(C)\le 0$. It is straightforward to see that for the inequality (a) to become an equality it is necessary and sufficient that $P^k(C)=1$ for all values $k$ of the answer message. Clearly, in that case, we can define an equivalent message $V'(C)$ that takes a single value $s$ so that $P^s(C)=1$.

It appears reasonable to define a perfect answer $V^*(C)$ to an ideal question $C$ as a message taking a single value $s$ such that $P^s(C)=1$. Note that while such a definition may sound a bit strange (since the answer takes a single value), it makes good sense if one keeps in mind that ideal question are auxiliary constructions and any answer to an ideal question should really be thought of a ``part'' of an answer to come complete question. We can again state the result obtained above as a lemma.

\begin{lemma}
Let $C$ be an ideal question and $V(C)$ an answer to it. Then $Y(\Omega,C,P,V(C))\le G(\Omega,C,P)$ with the inequality being tight if and only if the answer $V(C)$ is perfect.
\label{l:ans-fr}
\end{lemma}

Finally, let $\bC=\{C_1,\dotsc, C_r\}$ be a complete question and $\bC'=\{C_1,\dotsc, C_l\}$ where $l<r$ be an incomplete question. We define a perfect answer $V^*(\bC')$ to an incomplete question as a message taking values in the set $\{s_1,\dotsc, s_l\}$ such that $P^j(C_j)=1$ for $j=1,\dotsc, l$. As usual, $\hat C'=\cup_{j=1} C_j$.
Then it is straightforward to prove a result analogous to that of Lemmas~\ref{l:ans-mc}~and~\ref{l:ans-fr}.

 \begin{lemma}
  Suppose $\bC'$ is an incomplete question and $V(\bC')$ is an answer to it. Then $Y(\Omega,\bC',P,V(\bC'))\le G(\Omega,\bC',P)$ with the inequality being tight if and only if the answer $V(\bC')$ is perfect.
 \label{l:ans-mixed}
 \end{lemma}

 {\bf Proof:}  We can write the depth functional for $V(\bC')$ as follows.
 \begin{align*}
 & Y(\Omega,\bC',P,V(\bC')) = \sum_{k=1}^m \hat q_k \frac{\sum_{j=1}^l u(C_j)P^k(C_j)\log \frac{P^k(C_j)}{P(C_j)}}{P^k(\hat C')}\\
  &\overset{(a)}{=} \sum_{k=1}^m v_k \frac{\sum_{j=1}^l u(C_j)P^k(C_j)\log \frac{P^k(C_j)}{P(C_j)}}{P(\hat C')}\\
  &= \frac{1}{P(\hat C')}\sum_{k=1}^m \sum_{j=1}^l v_k u(C_j)P^k(C_j)\log P^k(C_j)\\ &- \frac{1}{ P(\hat C')} \sum_{k=1}^m \sum_{j=1}^l v_k u(C_j)P^k(C_j)\log P(C_j)\\
  &\overset{(b)}{=} \frac{1}{P(\hat C')}\sum_{k=1}^m \sum_{j=1}^l v_k u(C_j)P^k(C_j)\log P^k(C_j) - \frac{1}{P(\hat C')} \sum_{j=1}^l u(C_j)P(C_j)\log P(C_j)\\
  &= \frac{1}{P(\hat C')}\sum_{k=1}^m \sum_{j=1}^r v_k u(C_j)P^k(C_j)\log P^k(C_j) + G(\Omega,\bC',P)\overset{(c)}{\le} G(\Omega,\bC',P),
 \end{align*}
 where (a) follows from (\ref{eq:hat-qk}), (b) follows from the consistency with prior condition (\ref{eq:cons-mc}) for the complete question $\bC$, and (c) follows from the inequality $\log P^k(C_j)\le 0$. Using the same arguments as those employed for the proof of Lemma~\ref{l:ans-mc} we arrive at the statement of this lemma. \qed

 One can summarize the main result of this section by saying that, for any question type, the depth of any corresponding answer cannot exceed the difficulty of the question. Moreover, the answer depth can only be equal to the question difficulty in case the answer is perfect, i.e. the answer fully resolves the uncertainty associated with the question and does so with certainty. The number of different values of a perfect answer is always equal to the number of subsets in the question. While incomplete questions -- including ideal questions -- are just useful auxiliary constructs, the same basic property holds for them as well, at least for the isotropic knowledge structure model considered in the present article.

 \section{\label{s:quasi-perfect}Quasi-perfect answers}
 Let the question $\bC=\{C_1,\dotsc, C_r\}$ be complete and let $V(\bC)$ be an answer to $\bC$. If $V(\bC)$ is perfect, its depth $Y(\Omega, \bC, P, V(\bC))$ is equal to the difficulty $G(\Omega, \bC, P)$ of $\bC$ as Lemma~\ref{l:ans-mc} states. Here we would like to consider some simple classes of imperfect answers. To make the form of an imperfect answer more specific let us assume such as answer to resemble a  perfect one in that the number of possible values it can take is equal to $r$ and each message $s_k$, $k=1,\dotsc, r$ expresses a degree of preference towards the subset $C_k$. Let $e_k$ be the error probability associated with $s_k$, i.e $e_k=P^k(\bar C_k)$, where $\bar C_k=\Omega\setminus C_k$. Let us also make the additional assumption that the error associated with $s_k$ is ``proportionally distributed'' between sets $C_j$ $j\ne k$, i.e. $P^k(C_j)=\frac{e_k P(C_j)}{P(\bar C_k)}=\frac{e_k P(C_j)}{1-P(C_k)}$. Obviously, both of these assumptions can be stated in the following way.
 \begin{equation*}
 P^k=(1-e_k)P_{C_k}+\sum_{j\ne k} \frac{e_k P(C_j)}{1-P(C_k)}P_{C_j},
 \end{equation*}
 implying that the coefficients $p_{kj}$ in (\ref{eq:Pk-mc}) have the form
 \begin{equation}
 p_{kj}=\left(1-\frac{e_k}{1-P(C_k)}\right)\delta_{kj}+\frac{e_k P(C_j)}{1-P(C_k)}
 \label{eq:pkj-imp}
 \end{equation}

 To further simplify the analysis and provide more concise description of errors associated with imperfect answers we make a further assumption: that the error probability $e_k$ constitutes the same fraction of $P(\bar C_k)$ for all values of $k$, i.e. $e_k=\alpha (1-P(C_k))$, $k=1,\dotsc, r$, where $0\le \alpha\le 1$. Under this assumption, the error associated with the answer $V(\bC)$ that we will denote by $V_{\alpha}(\bC)$ is fully described by a single parameter $\alpha$. The coefficients $p_{kj}$ in (\ref{eq:pkj-imp}) become
 \begin{equation}
 p_{kj}=(1-\alpha)\delta_{kj}+\alpha P(C_j),
 \label{eq:pkj-alpha}
 \end{equation}
 and the updated measure $P^k$ becomes simply
 \begin{equation}
 P^k = \alpha P + (1-\alpha) P_{C_k}.
 \label{eq:Pk-alpha}
 \end{equation}
 We see that, for $\alpha=0$, measure $P^k$ turns into the conditional measure $P_{C_k}$ making the answer perfect, and for $\alpha=1$ each measure $P^k$ becomes the original measure $P$ thus rendering the answer $V_{\alpha}(\bC)$ empty, i.e. possessing vanishing depth.

 Substituting (\ref{eq:Pk-alpha}) into the general expression for the answer depth and using the fact that in this case $v_k=P(C_k)$, $k=1,\dotsc, r$, we can obtain
 \begin{equation}
 \begin{split}
 Y(\Omega,\bC,P,V_{\alpha}(\bC))&=\sum_{k=1}^r u(C_k)P(C_k)(1-\alpha +\alpha P(C_k))\log \frac{1-\alpha +\alpha P(C_k)}{P(C_k)}\\
 &+ \alpha \log \alpha \sum_{k=1}^r u(C_k)P(C_k)(1-P(C_k)),
 \end{split}
 \label{eq:Y-qp}
 \end{equation}
 It is easy to see that the expression (\ref{eq:Y-qp}) becomes $G(\Omega,\bC,P)$ for $\alpha=0$ and vanishes for $\alpha=1$.

 In the following we will call answers characterized by updated measures of the form (\ref{eq:Pk-alpha})
 and depth functionals given by (\ref{eq:Y-qp}) the {\it quasi-perfect} answers. Their advantage is that they allow to smoothly interpolate between perfect and empty answers using just a single parameter $\alpha$ taking values on the interval $[0,1]$.

 Substituting (\ref{eq:pkj-alpha}) into the consistency condition (\ref{eq:cons-mc}) it is easy to see that for quasi-perfect answers
 \begin{equation}
 v_j = P(C_j),
 \label{eq:v-alpha}
 \end{equation}
 for $j=1,\dotsc, r$, regardless of the value of error probability $\alpha$.

 \section{\label{s:relations}Relationships between questions and answers}
 Given two complete questions $\bC'$ and $\bC''$ the {\it pseudoenergy overlap} $J(\Omega,(\bC';\bC''),P)$ was defined in \cite{part1} as
 \begin{equation}
 J(\Omega,(\bC';\bC''),P)=G(\Omega,\bC',P)+G(\Omega,\bC'',P)-G(\Omega,\bC'\cap \bC'',P),
 \label{eq:overlap}
 \end{equation}
 which can easily be seen to have the following form
 \begin{equation}
 J(\Omega,(\bC';\bC''),P) = \sum_{i=1}^{r'} \sum_{j=1}^{r''} u(C'_i\cap C''_j)P(C'_i\cap C''_j)
 \log \frac{P(C'_i\cap C''_j)}{P(C'_i) P(C''_j)}.
 \label{eq:overlap(uP)}
 \end{equation}

 It was also shown that the pseudoenergy overlap can be interpreted as the reduction of difficulty of question $\bC''$ due to the knowledge of a perfect answer $V^*(\bC')$ to question $\bC'$.
 \begin{equation}
 G(\Omega,\bC'',V^*(\bC')) =  G(\Omega, \bC'',P)-J(\Omega, (\bC';\bC''), P),
 \label{eq:G=G-J}
 \end{equation}
 where the {\it conditional difficulty} $G(\Omega,\bC'',V(\bC'))$ is defined (for any answer $V(\bC')$ to question $\bC'$) as
 \begin{equation}
 G(\Omega,\bC'',V(\bC'))= \sum_{k=1}^{m'} \Pr(V(\bC')=s_k) G(\Omega, \bC'', P'^k).
 \label{eq:cond-diff}
 \end{equation}

 It would be interesting to find out how the relation (\ref{eq:G=G-J}) generalizes for the case of an arbitrary answer to question $\bC'$. Clearly, since a reception of value $s'_k$ of $V(C')$ updates the measure $P$ to $P'^k$, the difficulty of $\bC''$ given $V(C')=s'_k$ is equal to
\begin{equation*}
\begin{split}
G(\Omega, \bC'', P'^k) &= -\sum_{j=1}^{r''} u(C''_j)P'^k(C''_j)\log P'^k(C''_j)\\
                       &= -\sum_{j=1}^{r''} \sum_{l=1}^{r'} u(C'_l\cap C''_j)P'^k(C'_l\cap C''_j)\log P'^k(C''_j),
\end{split}
\end{equation*}
and therefore the
overall (expected) difficulty $G(\Omega,\bC'',V(\bC'))$ of question $\bC''$ given an answer $V(\bC')$ to $\bC'$ can be written -- denoting $\Pr(V(\bC')=s'_k)$ by $v'_k$ -- as
\begin{equation}
\label{eq:GC''V'}
\begin{split}
& G(\Omega,\bC'',V(\bC')) \equiv \sum_{k=1}^{m'} v'_k G(\Omega, \bC'', P'^k)
                       = -\sum_{k=1}^{m'} v'_k \sum_{j=1}^{r''} \sum_{l=1}^{r'} u(C'_l\cap  C''_j)P'^k(C'_l\cap C''_j)\log P'^k(C''_j)\\
                       &= \sum_{k=1}^{m'} v'_k \left(  \sum_{j=1}^{r''} \sum_{l=1}^{r'} u(C'_l\cap C''_j)P'^k(C'_l\cap C''_j)\log P'^k(C''_j)\right. \\ &+ \left. \sum_{j=1}^{r''} \sum_{l=1}^{r'} u(C'_l\cap C''_j)P'^k(C'_l\cap C''_j)\log P(C''_j) -  \sum_{j=1}^{r''} \sum_{l=1}^{r'} u(C'_l\cap C''_j)P'^k(C'_l\cap C''_j)\log P(C''_j)\right)\\
                       &= -\sum_{k=1}^{m'} v'_k \sum_{j=1}^{r''} \sum_{l=1}^{r'} u(C'_l\cap C''_j)P'^k(C'_l\cap C''_j)\log \frac{P'^k(C''_j)}{P(C''_j)}\\
                       &- \sum_{j=1}^{r''} \sum_{l=1}^{r'} u(C'_l\cap C''_j)P'^k(C'_l\cap C''_j)\log P(C''_j)\\
                       &= G(\Omega,\bC'',P)- \sum_{k=1}^{m'} v'_k \sum_{j=1}^{r''} \sum_{l=1}^{r'} u(C'_l\cap C''_j)P'^k(C'_l\cap C''_j)\log \frac{P'^k(C''_j)}{P(C''_j)}.
\end{split}
\end{equation}
We see from (\ref{eq:GC''V'}) that the conditional difficulty of $\bC''$ can be represented as a difference of the standard (unconditional) difficulty and another expression that can be appropriately denoted $Y(\Omega,\bC'',P,V(\bC'))$ and called the {\it relative depth} of the answer $V(\bC')$
with respect to question $\bC''$:
\begin{equation}
G(\Omega,\bC'',V(\bC')) = G(\Omega,\bC'',P) - Y(\Omega,\bC'',P,V(\bC')),
\label{eq:Gc=G-Yrel}
\end{equation}
where the relative depth $Y(\Omega,\bC'',P,V(\bC'))$ is given by
\begin{equation}
Y(\Omega,\bC'',P,V(\bC')) = \sum_{k=1}^{m'} v'_k \sum_{j=1}^{r''} \sum_{l=1}^{r'} u(C'_l\cap C''_j)P'^k(C'_l\cap C''_j)\log \frac{P'^k(C''_j)}{P(C''_j)}.
\label{eq:Yrel}
\end{equation}
Using the expression (\ref{eq:Pk-mc}) for the updated measures $P'^k$ we find that
\begin{equation}
P'^k(C'_l\cap C''_j)=p_{kl}\frac{P(C'_l\cap C''_j)}{P(C'_l)}
\label{eq:P'k(C'capC'')}
\end{equation}
and
\begin{equation}
P'^k(C''_j)=\sum_{l=1}^{r'} p_{kl}\frac{P(C'_l\cap C''_j)}{P(C'_l)},
\label{eq:P'k(C'')}
\end{equation}
and, substituting (\ref{eq:P'k(C'capC'')}) and (\ref{eq:P'k(C'')}) into (\ref{eq:Yrel}) we obtain for the relative depth:
\begin{equation}
Y(\Omega,\bC'',P,V(\bC')) = \sum_{k=1}^{m'} v'_k \sum_{l=1}^{r'} \sum_{j=1}^{r''} u(C'_l\cap C''_j)p'_{kl}\cdot \frac{P(C'_l\cap C''_j)}{P(C'_l)} \log \sum_{i=1}^{r'}p'_{ki}\cdot \frac{P(C'_i\cap C''_j)}{P(C'_i)\cdot P(C''_j)}.
\label{eq:Yrel-res}
\end{equation}

We can summarize the result just obtained as a lemma.
\begin{lemma}
Let $\bC'$ and $\bC''$ be two arbitrary complete questions on $\Omega$ and let $V(\bC')$ be an answer to $\bC'$. Then the conditional difficulty of $\bC''$ given the answer $V(\bC')$ can be found as
$$G(\Omega,\bC'',V(\bC')) = G(\Omega,\bC'',P) - Y(\Omega,\bC'',P,V(\bC')), $$
where the relative depth of $V(\bC')$ is given by the expression (\ref{eq:Yrel-res}).
\label{l:cond-diff}
\end{lemma}

Suppose now that $V^*(\bC')$ is a perfect answer to $\bC'$ which implies that $m'=r'$ and $p'_{kl}=\delta_{kl}$. Substituting this into (\ref{eq:Yrel-res}) and performing the sum over $k$ while making use of the answer consistency condition (\ref{eq:cons-mc}) we obtain
\begin{equation}
Y(\Omega,\bC'',P,V^*(\bC'))= \sum_{l=1}^{r'} \sum_{j=1}^{r''} u(C'_l\cap C''_j)P(C'_l\cap C''_j)
\log \frac{P(C'_l\cap C''_j)}{P(C'_l) P(C''_j)},
\label{eq:Yrel-perf}
\end{equation}
which coincides with the expression (\ref{eq:overlap(uP)}) for the pseudoenergy overlap between questions $\bC'$ and $\bC''$. We thus recover the result (\ref{eq:G=G-J}) obtained in \cite{part1}.

Let now $V_{\alpha}(\bC')$ be a quasi-perfect answer to question $\bC'$ characterized by error probability $\alpha$. Substituting expressions (\ref{eq:pkj-alpha}) and (\ref{eq:v-alpha}) into (\ref{eq:Yrel-res}) we obtain, after some straightforward algebra
\begin{equation}
\begin{aligned}
& Y(\Omega,\bC'',P,V_{\alpha}(\bC'))=(1-\alpha)\sum_{l=1}^{r'} \sum_{j=1}^{r''} u(C'_l\cap C''_j)P(C'_l\cap C''_j)\log \left[(1-\alpha)\frac{P(C'_l\cap C''_j)}{P(C'_l) P(C''_j)}+\alpha\right]\\
&+\alpha \sum_{l=1}^{r'} \sum_{j=1}^{r''} u(C'_l\cap C''_j)P(C'_l\cap C''_j)\sum_{k=1}^{r'} P(C'_k)\log \left[(1-\alpha)\frac{P(C'_k\cap C''_j)}{P(C'_k) P(C''_j)}+\alpha\right]
\end{aligned}
\label{eq:Yrel-alpha}
\end{equation}
It is easy to see that for $\alpha=0$ (\ref{eq:Yrel-alpha}) reduces to (\ref{eq:Yrel-perf}) which is the overlap between questions $\bC'$ and $\bC''$, and for $\bC''$ coinciding with $\bC'$ the relative depth (\ref{eq:Yrel-alpha}) becomes the depth $Y(\Omega,\bC',P,V_{\alpha}(\bC'))$ (given by expression (\ref{eq:Y-qp})) of quasi-perfect answer to $\bC'$ characterized by the same value of error probability $\alpha$. To see that, it is sufficient to set $C''_j=C'_j$ (and hence $P(C'_l\cap C''_j)=\delta_{lj}P(C'_l)$) in (\ref{eq:Yrel-alpha}) and make use of the (obvious) identity $\sum_{k\ne j}P(C'_k)=1-P(C'_j)$.

\section{\label{s:sources}Information source models}
The question difficulty functional describes the source knowledge structure by specifying the amount of pseudoenergy associated with any question $\bC$. The answer depth, on the other hand, quantifies the pseudoenergy associated with any answer of the source to question $\bC$, with more accurate answers carrying larger amounts of pseudoenergy. The next logical question is how accurately can the given information source answer the specific question $\bC$. This question can be restated by asking what value of answer depth the source is able to provide in response to $\bC$. It would be very natural to assume that this depth should be a function of the difficulty of question $\bC$. This assumption essentially implies that the corresponding question difficulty faithfully characterizes the source knowledge structure. Finding this assumption to be wrong can be looked upon as an indication of the agent's failure to identify some essential features of the source knowledge structure (like, for example, its anisotropy). We formulate this assumption as a hypothesis.

{\bf Hypothesis S1}. For the given information source and any question $\bC$, the corresponding answer depth is a function of the question difficulty:
$$Y(\Omega,\bC,P,V(\bC))=h(G(\Omega,\bC,P)),$$
where $h$: $\bR_+\rightarrow \bR_+$ is a function of a single argument.

Note that Hypothesis S1 can be thought of as that of existence of some knowledge structure (quite possibly a fairly complicated one) for the given information source. Put slightly differently, it states that the set of all possible questions can be represented as a partially ordered set with respect to the source's ability to answer them accurately. If this is indeed true, symmetry and consistency considerations can be invoked to find the specific form of the question difficulty and answer depth functionals.

The specific shape of function $h(\cdot)$ can be determined by experimentation: one would generally have to assume some reasonable overall shape and then use sample questions and the source's answers to estimate parameters. (The procedure is similar to that applied, for instance, to fitting regression models.) If a particular (simple) model for $h(\cdot)$ is found inadequate, more elaborate models can be employed, until a good stable fit can be established. We briefly discuss some possible models next.

\subsection{Possible source models}
The overall shape of function $h(\cdot)$ is, in general, arbitrary, but it would be reasonable to make some initial assumptions on the grounds of general experience. In particular, it would be sensible to assume that the function $h(\cdot)$ possesses one or more of the following properties: (i) non-decreasing; (ii) continuous; (iii) bounded from above. Property (i) simply implies that the source can produce at least as much depth when answering a more difficult question. Property (ii) means that if two questions are close in difficulty the source will produce answers of close depth. And property (iii) implies existence of a well-defined {\it source capacity} -- the highest answer depth the source is capable of.
Let us now take a look at some simple models satisfying these properties.

 {\bf Simple capacity model}

 In this model, the information source is characterized by a single parameter that the source (pseudoenergy) capacity which we denote by $Y_s$. Under this model, the source can provide perfect answers to questions whose difficulty does not exceed $Y_s$ and, for questions with difficulty exceeding $Y_s$, the error probabilities increase in such a way that the depth of the corresponding answer stays equal to $Y_s$. Put slightly differently, the source provides answers whose depth is constant unless the question is too easy for the source in which case the depth of the answer is limited by the difficulty of the question. Formally speaking, the function $h(x)$ for this model takes the following form.
 \begin{equation}
 h(x)=\begin{cases} x & \mbox{if}\; x\le Y_s \\
                   Y_s & \mbox{if}\; x>Y_s.
 \end{cases}
 \label{eq:Y(G)-cap-simple}
 \end{equation}

 {\bf Modified capacity models}

 The main drawback of the simple capacity model described above is that the information source is postulated to provide perfect answers to questions whose difficulty is below the source's capacity. On the other hand, in many situations, it is reasonable to expect that a source will make some error answering even simple questions. The modified capacity models' goal is allow for finite error probabilities for answers to questions with difficulties below the source capacity. This model depends on more than one two parameter: besides the capacity $Y_s$, there is also a parameter describing way function $h(\cdot)$ approaches its maximum value $Y_s$. The simplest such models is the linear modified capacity model described by
 \begin{equation}
 h(x)=\begin{cases} bx & \mbox{if}\; x\le \frac{Y_s}{b} \\
                   Y_s & \mbox{if}\; x>\frac{Y_s}{b}.
 \end{cases}
 \label{eq:Y(G)-cap-mod}
 \end{equation}
 where $b\le 1$ is the second parameter. Under this model, the source makes errors even on questions with difficulties below the capacity with error probabilities gradually increasing with question difficulties. Once the question difficulty exceeds the capacity of the source, the corresponding answer depth stays equal to the capacity $Y_s$.

 The linear modified capacity model can be naturally generalized to a polynomial modified capacity model in which the function $h(\cdot)$ approaches its maximum value according to a polynomial law. To describe it, let $p_q(x)=a_0+a_1x+\dotsc + a_qx^q$ be an order $q$ polynomial and let $x^*_q$ be the smallest positive root of the equation $p_q(x)-Y_s=0$. Then the {\it polynomial modified capacity model} has the form
 \begin{equation}
 h(x)=\begin{cases} p_q(x) & \mbox{if}\; x\le x^*_q \\
                   Y_s & \mbox{if}\; x>x^*_q.
 \end{cases}
 \label{eq:Y(G)-cap-mod-poly}
 \end{equation}
 Demanding that $h(0)=0$ and $h(x)\le x$ for all $x\ge 0$ leads to $a_0=0$ and $0\le a_1\le 1$. For $q=2$, the polynomial modified capacity model (\ref{eq:Y(G)-cap-mod-poly}) reduces to the {\it quadratic modified capacity} model that has the form
 \begin{equation}
 h(x)=\begin{cases} bx+cx^2 & \mbox{if}\; x\le x^*_2 \\
                   Y_s & \mbox{if}\; x>x^*_2,
 \end{cases}
 \label{eq:Y(G)-cap-mod-quad}
 \end{equation}
 where $0< b\le 1$ and (assuming $c<0$) $|c|\le \frac{b^2}{4Y_s}$; $x^*_2=\frac{b}{2|c|}-\frac{\sqrt{b^2-4|c|Y_s}}{2|c|}$.

 Another simple model that belongs to the class of modified capacity models is the {\it exponential modified capacity model}
 \begin{equation}
 h(x)= Y_s(1-e^{-\theta x})
 \label{eq:Y(G)-cap-mod-exp}
 \end{equation}
 that depends on two parameters: capacity $Y_s$ and $0<\theta\le \frac{1}{Y_s}$ that controls the speed with which the function $h(x)$ approaches its upper bound $Y_s$. One of the advantages of the exponential model (\ref{eq:Y(G)-cap-mod-exp}) is that it's described by a single analytical function that allows to avoid binary variables in the corresponding estimation problem discussed in the next section.

\section{\label{s:estimation}Estimation of pseudotemperature and source model parameters}
In this section, just like in rest of this article and \cite{part1}, we assume that the linear isotropic model of the source knowledge structure holds.
First, let us note that both question difficulty and answer depth functionals are linear in  $u(\o)$ and therefore multiplying $u(\o)$ by any constant would result in both difficulty and depth being multiplied by the same constant without changing any of the coefficients $p_{kj}$, $k=1,\dotsc, m$, $j=1,\dotsc, r$ and, therefore, answer error probabilities. This means that the function $u(\o)$ is really defined up to a single multiplicative constant the choice of which is equivalent to a choice of units in which $u(\o)$ (and the difficulty/depth functionals) are measured. We use two different conventions that turn out to be convenient.
 \begin{itemize}
 \item The normalized $u(\o)$ convention in which $\int_{\Omega} u(\o)dP(\o)=1$ for every information source. This convention is convenient because if $u(\o)\equiv 1$, the  difficulty of question $\bC$ reduces to Shannon entropy of the distribution $P(\bC)=(P(C_1),\dotsc, P(C_r))$. In a sense, this allows for measuring pseudoenergy in units commensurate with standard information bits.

 \item The unit source capacity convention in which the units of $u(\o)$ are chosen in such a way
 that, for each information source, the source capacity (assuming it exists) is unity: $Y_s=1$. This convention is useful for comparing different information sources to each other. Indeed, in this case, functions $u(\o)$ for any two sources can be directly compared to each other showing clearly the relative degree of ``expertise'' of each source in various regions of $\Omega$ and also giving a sense of ``absolute'' quality of each source.
 \end{itemize}

 If the function $u(\o)$ is known Theorem~\ref{th:G} gives (for the given measure $P$) the difficulty of any question $\bC$. Then, for any answer $V(\bC)$ to $\bC$, the knowledge of updated measures $P^k$ allows one to find the depth of $V(\bC)$. On the other hand, a given source model $Y=h(G)$ lets one {\it predict} the depth of the source's answer to any question before measures $P^k$ can be estimated. Thus in order to be able to predict the depth of source's answer to various questions one needs to know the function $u(\o)$ and the source model described by the function $h(\cdot)$. Since these functions cannot be directly measured or observed, the only way to know them in any application is to estimate them from the source's performance on a certain set of sample questions.

 Let $\bD=\{D_1,\dotsc, D_{N_d} \}$ be a partition of $\Omega$ that to be used for discretizing the weight function $u(\o)$: we assume that $u(\o)$ takes a constant value equal to $u_i$ on subset $D_i$. Let $w_i=P(D_i)$ and let ${\cal N}_i \subset \{1,\dotsc, N_d\}$ be index set of subsets in $\bD$ that are immediate neighbors (i.e. have a common boundary with) of subset $D_i$. We assume that the partition $\bD$ is sufficiently fine so that any partition $\bC$ used for estimating $u(\o)$ can be considered a coarsening of $\bD$.

 Further, let $\bC_1,\dotsc, \bC_K$ be a set of questions that the source has answered and its answers have been compared with actual outcomes in $\Omega$.
 Let us denote by $G_1,\dotsc, G_K$ be difficulties of these questions and let $Y_1,\dotsc, Y_K$ be the corresponding answer depth values that were computed using the estimated error probabilities \footnote{
 For the sake of simplicity, we assume that the answers of the source are quasi-perfect with the corresponding (estimated) error probabilities being equal to $\alpha_1, \dotsc, \alpha_K$, respectively.}.

 Let us introduce the notation $z_i=|Y_i-h(G_i)|$, $i=1,\dotsc, K$ where the function $h(\cdot)$ is given by the suitable information source model. The quantities $z_i$ measure the absolute values of deviations of the empirical data from the chosen source model, with vanishing values of all variables $z_i$ corresponding to a perfect fit. In addition to minimizing the sum of the deviations (i.e. maximizing the fit), it makes sense to demand that the quantities $u_j$, $j=1,\dotsc, N_d$, describe a reasonably smooth function $u(\o)$. This can be achieved, for instance, by putting an upper bound on the gradient of $u(\o)$ or, equivalently, by putting a corresponding term in the objective function. To make it more precise, let $N(\bD)$ be the set of neighbors in the partition $\bD$ and let $U$ be the desired upper bound on the difference of two values of $u$ on neighboring sets of partition $\bD$. Then, if the capacity model $h(\cdot)$ is postulated, the following formulation of the estimation problem for the function $u(\o)$ and the parameters of model $h(\cdot)$ is obtained.

 \begin{equation}
\begin{aligned}
& \text{minimize}
& & \sum_{i=1}^{K} z_i + \lambda U \\
& \text{subject to}
& & Y_i-h(G_i) \le z_i,\; i=1,\dotsc, K\\
& & & h(G_i)-Y_i\le z_i,\; i=1,\dotsc, K\\
& & & u_j-u_k \le U,\; (j,k)\in N(\bD)\\
& & & u_k-u_j \le U,\; (j,k)\in N(\bD)
\end{aligned}
\label{eq:uest-gen}
\end{equation}
The decision variables in (\ref{eq:uest-gen}) are $z_i$, $u_j$, $j=1,\dotsc, N_d$ and the  parameters of function $h(\cdot)$. $\lambda$ is a parameter that controls the trade-off between the objective of maximizing the fit and that of maximizing smoothness of $u(\o)$ (understood as minimizing the maximum gradient of $u(\o)$). The difficulties $G_i$, $i=1,\dotsc, K$ are expressed via the decision variables as follows.
\begin{equation}
G_i=-\sum_{j=1}^{r_i} \left(\sum_{\{l: D_l\subset C_j\}} u_lw_l \right) \log P(C_j)
\label{eq:Gi}
\end{equation}
For the values of the depth functional for the corresponding answers, let us assume, for simplicity, that the answers are quasi-perfect implying that their errors can be characterized with a single probability $\alpha_i$, $i=1,\dotsc, K$. Then the depth $Y_i$ can be written as
\begin{equation}
\begin{split}
 Y_i &=\sum_{j=1}^{r_i} (1-\alpha_i +\alpha_i P(C_j))\log \frac{1-\alpha_i +\alpha_i P(C_j)}{P(C_j)} \left(\sum_{\{l: D_l\subset C_j\}} u_lw_l \right)\\
 &+ \alpha_i \log \alpha_i \sum_{j=1}^{r_i} P(C_j)\left(1-\sum_{\{l: D_l\subset C_j\}} u_lw_l \right),
 \end{split}
 \label{eq:Yi}
\end{equation}

Note that, in general, (\ref{eq:uest-gen}) is a potentially complex nonlinear optimization problem where nonlinearity is introduced by the function $h(\cdot)$. For the case of the simple capacity model, the problem (\ref{eq:uest-gen}) can be written as
\begin{equation}
\begin{aligned}
& \text{minimize}
& & \sum_{i=1}^{K} z_i + \lambda U \\
& \text{subject to}
& & Y_i-Y_s \le z_i + My_i,\; i=1,\dotsc, K \\
& & & Y_s-Y_i\le z_i +My_i,\; i=1,\dotsc, K \\
& & & G_i-Y_i\le z_i + M(1-y_i),\; i=1,\dotsc, K\\
& & & u_j-u_k \le U,\; (j,k)\in N(\bD)\\
& & & u_k-u_j \le U,\; (j,k)\in N(\bD)\\
& & & y_i\in \{0,1\},\; i=1,\dotsc, K
\end{aligned}
\label{eq:uest-scm}
\end{equation}
In this formulation, $M$ is a large number, $y_i$, $i=1,\dotsc K$, are auxiliary binary variables. The main decision variables in the formulation (\ref{eq:uest-scm}) are the values $u_j$, $j=1,\dotsc, N_d$, and the capacity value $Y_s$. Since both (\ref{eq:Gi}) and (\ref{eq:Yi}) are linear in the variables $u_l$, the optimization problem (\ref{eq:uest-scm}) is mixed-linear with $K$ binary variables and therefore can be solved efficiently at least for moderate values $K$ of sample questions used for estimating model parameter $Y_s$ and the (discretized) function $u(\o)$.

The formulation (\ref{eq:uest-scm}) can be modified easily from the simple to the modified capacity model. The resulting formulation is as follows.
\begin{equation}
\begin{aligned}
& \text{minimize}
& & \sum_{i=1}^{K} z_i + \lambda U \\
& \text{subject to}
& & Y_i-Y_s \le z_i + My_i,\; i=1,\dotsc, K \\
& & & Y_s-Y_i\le z_i +My_i, \; i=1,\dotsc, K \\
& & & bG_i-Y_i\le z_i + M(1-y_i), \; i=1,\dotsc, K\\
& & & u_j-u_k \le U,\; (j,k)\in N(\bD)\\
& & & u_k-u_j \le U,\; (j,k)\in N(\bD)\\
& & & y_i\in \{0,1\},\; i=1,\dotsc, K
\end{aligned}
\label{eq:uest-mcm}
\end{equation}
The additional decision variable in (\ref{eq:uest-mcm}) is $b\le 1$. The values $G_i$ and $Y_i$, $i=1,\dotsc K$ are given by expressions (\ref{eq:Gi}) and (\ref{eq:Yi}), respectively. The formulation
(\ref{eq:uest-mcm}) is,  just like (\ref{eq:uest-scm}), is a mixed-linear optimization problem with $K$ binary variables and thus can be solved efficiently at least for moderate values of the number $K$ of sample questions.

The formulation for the quadratic modified capacity model (\ref{eq:Y(G)-cap-mod-quad}) can be easily obtained from (\ref{eq:uest-mcm}) by replacing the constraints $bG_i-Y_i\le z_i + M(1-y_i)$, $i=1,\dotsc, K$ with $bG_i+cG_i^2-Y_i\le z_i + M(1-y_i)$, $i=1,\dotsc, K$. Recalling that $G_i$ is a linear function of the decision variables $u_l$, we see that the resulting problem is that of quadratic optimization with $K$ binary variables that enter the formulation in a linear fashion. Even thought such problems can't in general be solved as efficiently as mixed-linear optimization problems of equal size, they still can be solved to optimality for moderate values of parameters $K$ and $N_d$.

The modified exponential capacity model, as mentioned earlier, has the one advantage that the corresponding formulation of the estimation problem obviates the need for binary variables even though it becomes severely nonlinear:
\begin{equation}
\begin{aligned}
& \text{minimize}
& & \sum_{i=1}^{K} z_i + \lambda U \\
& \text{subject to}
& & Y_i-Y_s(1-e^{-\theta G_i}) \le z_i,\; i=1,\dotsc, K \\
& & & Y_s(1-e^{-\theta G_i})-Y_i\le z_i, \; i=1,\dotsc, K \\
& & & u_j-u_k \le U,\; (j,k)\in N(\bD)\\
& & & u_k-u_j \le U,\; (j,k)\in N(\bD)\\
& & & y_i\in \{0,1\},\; i=1,\dotsc, K
\end{aligned}
\label{eq:uest-exp}
\end{equation}
Besides the quantities $z_i$, $i=1,\dotsc, K$, $u_l$, $l=1,\dotsc, N_d$ and the source capacity $Y_s$, another decision variable is the parameter $0<\theta\le \frac{1}{Y_s}$.

It is worth noting that in estimation of the pseudotemperature function and model parameters, the error probabilities are themselves estimated values. That introduces obvious imprecision in estimation of pseudotemperature and source model parameters. In fact, one can think of the procedure described in this section as similar to point estimation of parameters in classical statistics. For more information about the pseudotemperature function, confidence intervals would be needed. The width of such confidence intervals would obviously depend on the precision with which error probabilities are known and therefore on the sample size used in error probability estimation. Practically, such confidence intervals may turn out to be sufficiently wide to effectively invalidate precise estimation of the shape of pseudotemperature function. The practical approach instead could be that of the hypothesis testing type: a null (default) hypothesis about the shape of the pseudotemperature function would be stated (i.e. that the pseudotemperature is constant or linear) and then tested using standard statistical methods.

Just like in probability estimation, expert opinion can be used for estimating pseudotemperature function.
Since pseudotemperature admits a simple intuitive interpretation (as local ``degree of difficulty'') experts should find it easy enough to give useful estimates of pseudotemperature. If, in addition, some data about observed source performance is available, it can be used in conjunction with expert estimates by, for instance, using expert estimate as a null hypothesis and using observed data for the purpose of testing it.

\section{\label{s:examples}Examples}
Let us revisit the example with a finite parameter space from \cite{part1}. The parameter space $\Omega$ consists of 8 elements, corresponding to green, yellow and red apples (denoted $GA$, $YA$ and $RA$, respectively), green, yellow and red pears (denoted $GPr$, $YPr$ and $RPr$), and yellow and red peaches (denoted $YPc$ and $RPc$). The elements are equiprobable so that $P(\cdot)=\frac{1}{8}$ for all $\o\in \Omega$. The function $u(\cdot)$ describes the relative difficulty of respective ideal questions. We set $u(GA)=u(GPr)=1$, $u(YPr)=u(RPr)=1.5$, and $u(YA)=u(RA)=u(YPc)=u(RPc)=2$. Normalizing the values of $u(\cdot)$ so that $\int_{\Omega} u(\o) dP(\o)=1$ we obtain $u(GA)=u(GPr)=\frac{8}{13}$, $u(YPr)=u(RPr)=\frac{12}{13}$ and $u(YA)=u(RA)=u(YPc)=u(RPc)=\frac{16}{13}$.

 Consider, as in \cite{part1}, the question {\it ``Is the fruit green or not?''}. Let $C_g=\{GA, GPr\}\subset \Omega$ be the subset consisting of all green fruit (apples and pears) and let $\overline C_g=\Omega\setminus C_g$ be the subset containing fruit of all other colors (red and yellow). The partition is $\bC_g=\{C_g, \overline C_g \}$. The values $u(\cdot)$ for the sets in this partition are $u(C_g)=\frac{8}{13}$ and $u(\overline C_g)=\frac{1}{3}\cdot \frac{12}{13}+\frac{2}{3}\cdot \frac{16}{13}=\frac{44}{39}$. The measures are $P(C_g)=\frac{1}{4}$ and $P(\overline C_g)=\frac{3}{4}$. The second similar question is {\it ``Is the fruit a peach or not?''}. The corresponding partition is $\bC_{Pc}=\{C_{Pc}, \overline C_{Pc}\}$ where $C_{Pc}=\{YPc, RPc\}$ and $\overline C_{Pc}=\Omega\setminus C_{Pc}$. The values of function $u(\cdot)$ on these subsets are $u(C_{Pc})=\frac{16}{13}$ and $u(\overline C_{Pc})=\frac{1}{3}\cdot \frac{8}{13}+\frac{1}{3}\cdot \frac{12}{13}+\frac{1}{3}\cdot \frac{16}{13}=\frac{12}{13}$. The measures are $P(C_{Pc})=\frac{1}{4}$ and  $P(\overline C_{Pc})=\frac{3}{4}$. Let $V_{\alpha}(\bC_g)$ and $V_{\alpha}(\bC_{Pc})$ be the corresponding quasi-perfect answers. The depth functionals of these answers can be computed using (\ref{eq:Y-qp}) as
 $$Y(\Omega, \bC_g, P, V_{\alpha}(\bC_g))= \frac{2}{13}\left( 1-\frac{3}{4}\alpha \right)\log (4-3\alpha)+ \frac{11}{13}\left( 1-\frac{1}{4}\alpha \right)\log \frac{4-\alpha}{3}+ \frac{15}{52}\alpha\log \alpha,$$
 and
 $$Y(\Omega, \bC_{Pc}, P, V_{\alpha}(\bC_{Pc}))= \frac{4}{13}\left( 1-\frac{3}{4}\alpha \right)\log (4-3\alpha)+ \frac{9}{13}\left( 1-\frac{1}{4}\alpha \right)\log \frac{4-\alpha}{3}+ \frac{21}{52}\alpha\log \alpha.$$

\begin{figure}[hbt]
\includegraphics[scale=0.38]{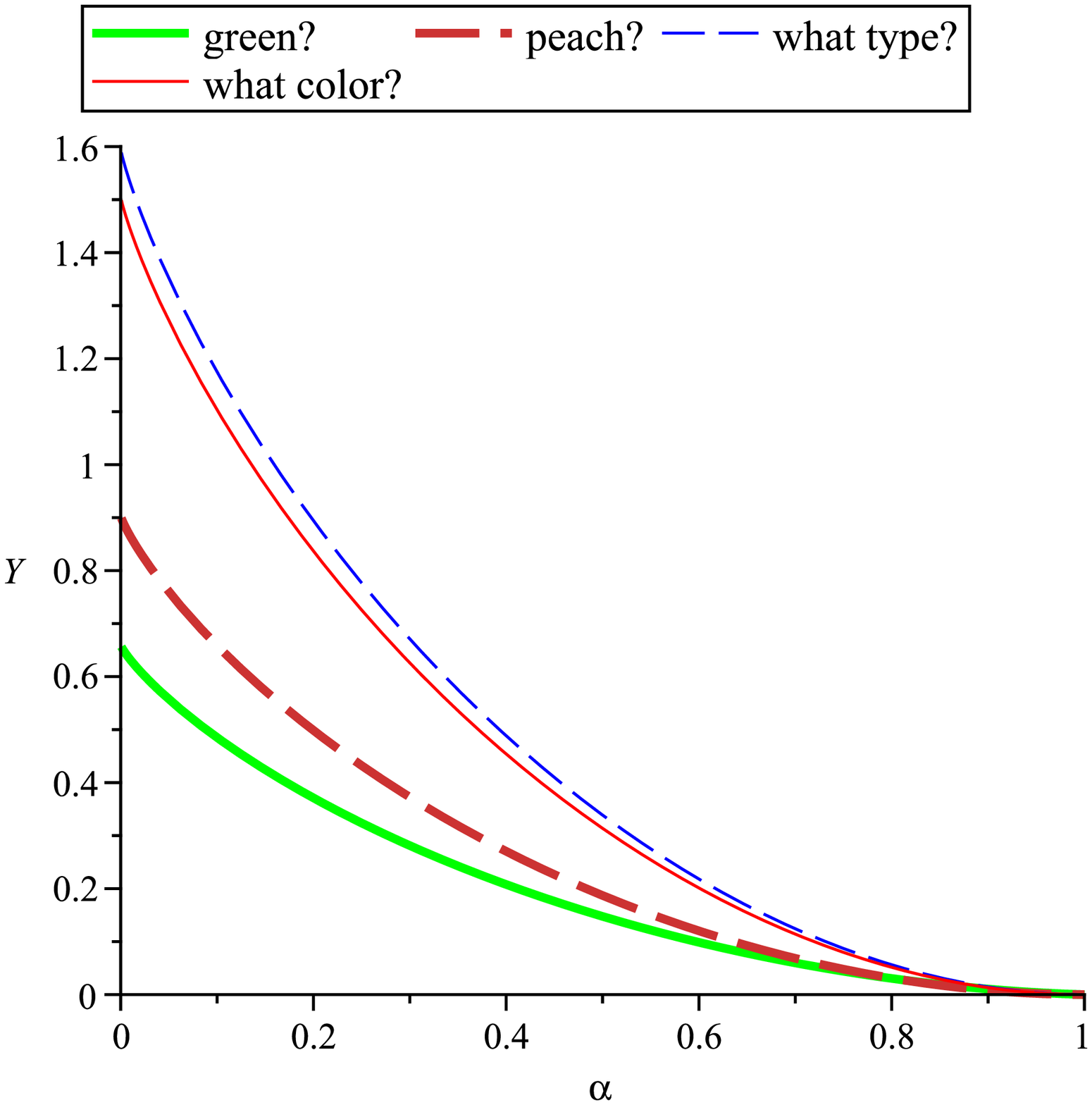}
\includegraphics[scale=0.38]{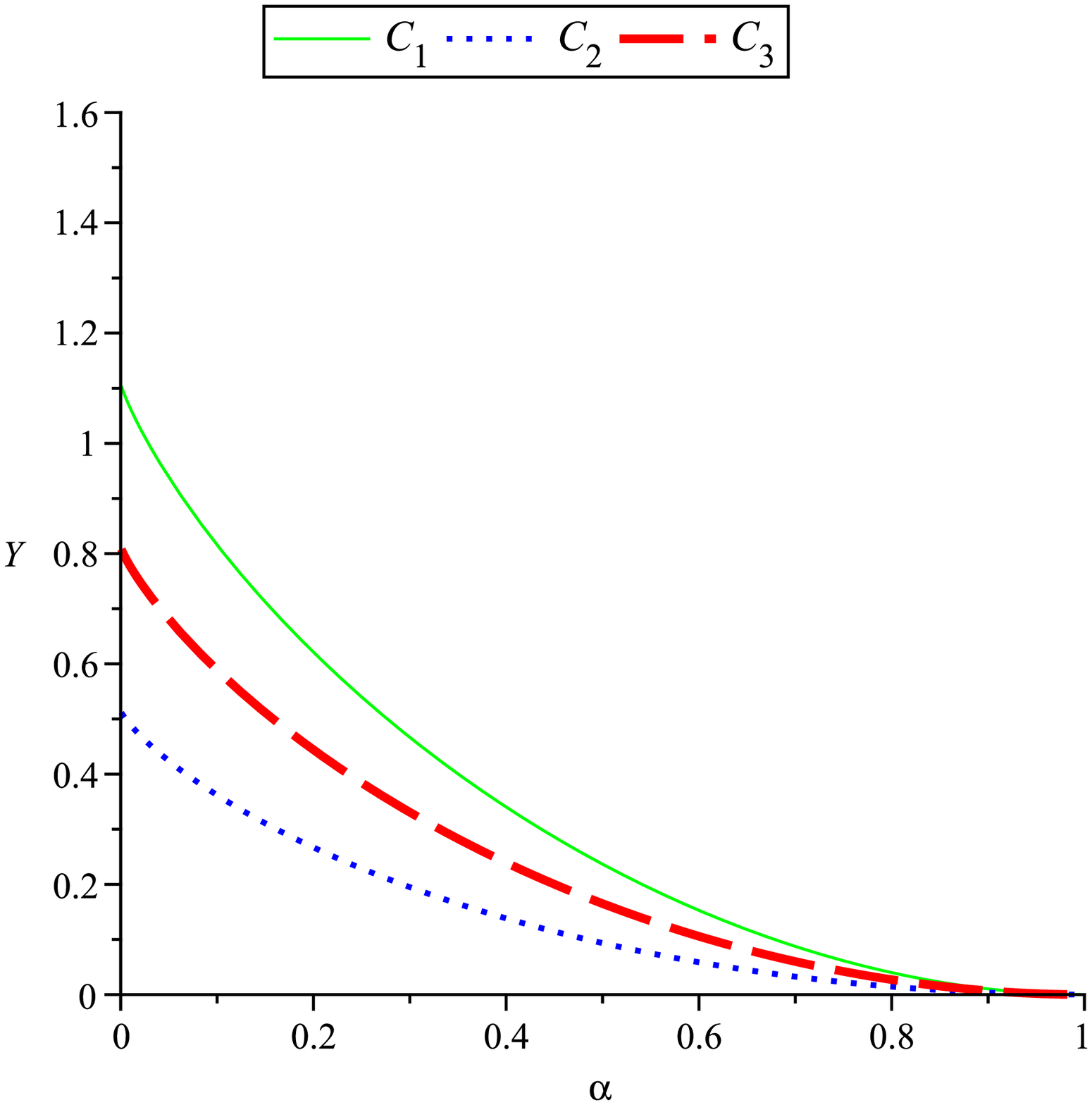}
\caption{ \label{f:depth}Answer depth as a function of $\alpha$ for quasi-perfect answers to questions on the finite parameter space (left) and infinite parameter space (right).  }
\end{figure}

Consider the question {\it ``What color is the given fruit?''} on one hand and {\it ``What type is the given fruit?''} on the other. The former question can be represented as the partition $\bC_c=\{C_g, C_y, C_r \}$ where $C_g=\{GA, GPr\}$, $C_y=\{YA, YPr, YPc\}$ and $C_r=\{RA, RPr, RPc\}$; the latter question can be identified with the partition $\bC_t=\{C_A, C_{Pr}, C_{Pc} \}$ where $C_A=\{GA, YA, RA\}$, $C_{Pr}=\{GPr, YPr, RPr\}$ and $C_{Pc}=\{YPc, RPc\}$. The values of $u(\cdot)$ on these subsets are $u(C_g)=\frac{8}{13}$, $u(C_y)=\frac{1}{3}\cdot \frac{12}{13}+\frac{2}{3}\cdot \frac{16}{13}=\frac{44}{39}$, $u(C_g)=u(C_y)=\frac{44}{39}$; $u(C_A)=\frac{1}{3}\cdot \frac{8}{13}+\frac{2}{3}\cdot \frac{16}{13}=\frac{40}{39}$, $u(C_{Pr})=\frac{1}{3}\cdot \frac{8}{13}+\frac{2}{3}\cdot \frac{12}{13}=\frac{32}{39}$, $u(C_{Pc})=\frac{16}{13}$. The measures are $P(C_g)=\frac{1}{4}$, $P(C_y)=\frac{3}{8}$, $P(C_r)=\frac{3}{8}$; $P(C_A)=P(C_{Pr})=\frac{3}{8}$, $P(C_{Pc})=\frac{1}{4}$. Let $V_{\alpha}(\bC_c)$ and $V_{\alpha}(\bC_t)$ be quasi-perfect answers to questions $\bC_c$ and $\bC_t$. The depth of these answers can be found using the expression (\ref{eq:Y-qp}). The results are (see Fig.~\ref{f:depth})
$$Y(\Omega, \bC_c, P, V_{\alpha}(\bC_c))= \frac{2}{13}\left( 1-\frac{3}{4}\alpha \right)\log (4-3\alpha)+ \frac{11}{13}\left( 1-\frac{5}{8}\alpha \right)\log \frac{8-5\alpha}{3}+ \frac{67}{104}\alpha\log \alpha,$$
and
$$Y(\Omega, \bC_t, P, V_{\alpha}(\bC_t))= \frac{4}{13}\left( 1-\frac{3}{4}\alpha \right)\log (4-3\alpha)+ \frac{9}{13}\left( 1-\frac{5}{8}\alpha \right)\log \frac{8-5\alpha}{3}+ \frac{69}{104}\alpha\log \alpha.$$

Let us consider the second example from \cite{part1}. The parameter space is $\Omega=[0,1]^2\subset \bR^2$. Let the pseudotemperature function be $u(\o)=\frac{3}{2}(\o_1^2+\o_2^2)$ (so that the hard questions are located towards the upper-right corner of $\Omega$). Consider the following three subsets of $\Omega$: $C_1=\{\o : \o_1\in [\frac{1}{2},1], \o_2\in [\frac{1}{2},1]\}$, $C_2=\{\o: \o_1\in [0,\frac{1}{2}], \o_2\in [0,\frac{1}{2}] \}$, $C_3=\{\o: \o_1\in [0,\frac{1}{2}], \o_2\in [\frac{1}{2},1] \}$ and let $\bC_i=\{C_i, \overline C_i\}$ for $i=1,2,3$ be three complete questions on $\Omega$. Let $V(\bC_i)$ be a quasi-perfect answer to question $\bC_i$, $i=1,2,3$ characterized by error probability $\alpha$. We can use the expression (\ref{eq:Y-qp}) to obtain the depth of these answers (see Fig.~\ref{f:depth} for an illustration).
$$Y(\Omega, \bC_1, P, V(\bC_1))=\frac{7}{16}\left(1-\frac{3}{4}\alpha\right)\log \left(4-3\alpha \right)+ \frac{9}{16}\left(1-\frac{1}{4}\alpha\right)\log\frac{4-\alpha}{3}+\frac{15}{32}\alpha \log \alpha,$$
$$Y(\Omega, \bC_2, P, V(\bC_2))=\frac{1}{16}\left(1-\frac{3}{4}\alpha\right)\log \left(4-3\alpha \right)+ \frac{15}{16}\left(1-\frac{1}{4}\alpha\right)\log\frac{4-\alpha}{3}+\frac{9}{32}\alpha \log \alpha,$$
and
$$Y(\Omega, \bC_3, P, V(\bC_3))=\frac{1}{4}\left(1-\frac{3}{4}\alpha\right)\log \left(4-3\alpha \right)+ \frac{3}{4}\left(1-\frac{1}{4}\alpha_1\right)\log\frac{4-\alpha}{3}+\frac{3}{8}\alpha \log \alpha.$$

Let us turn to relative depth of answers. Consider the above example again. The relative depth $Y(\Omega, \bC'',P, V_{\alpha}(\bC'))$ of a quasi-perfect answer $V_{\alpha}(\bC')$ with respect to question $\bC''$ can be readily computed using the expression (\ref{eq:Yrel-alpha}).
We obtain, for questions $\bC_1$ and $\bC_2$,
\begin{equation*}
\begin{split}
Y(\Omega, \bC_2, P, V_{\alpha}(\bC_1))&= \left(\frac{1}{2}-\frac{7}{32}\alpha\right)\log \left(\frac{4}{3}(1-\alpha)+\alpha \right)\\
 &+ \left(\frac{1}{2}+\frac{13}{64}\alpha\right)\log \left(\frac{8}{9}(1-\alpha)+\alpha \right) +\frac{1}{64}\alpha \log \alpha,
\end{split}
\end{equation*}

\begin{equation*}
\begin{split}
Y(\Omega, \bC_1, P, V_{\alpha}(\bC_2))&= \left(\frac{1}{2}-\frac{1}{32}\alpha\right)\log \left(\frac{4}{3}(1-\alpha)+\alpha \right) \\
&+ \left(\frac{1}{2}-\frac{5}{64}\alpha\right)\log \left(\frac{8}{9}(1-\alpha)+\alpha \right) +\frac{7}{64}\alpha \log \alpha.
\end{split}
\end{equation*}

\begin{figure}
\includegraphics[scale=0.48]{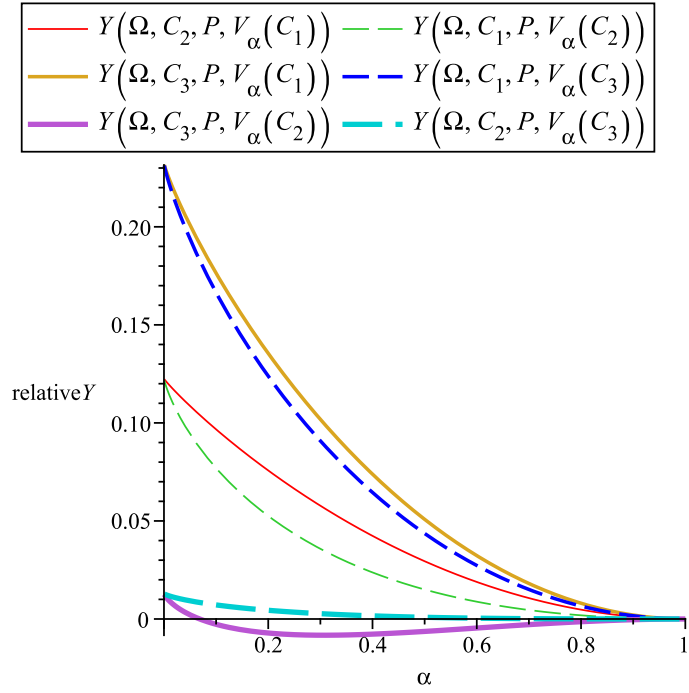}
\caption{\label{f:relative-Y}Relative depth of quasi-perfect answers as a function of $\alpha$.}
\end{figure}

Similar expressions obtain for the other two question pairs. The results are shown in~Fig.~\ref{f:relative-Y}.
We can see, in particular,
that the relative depth $Y(\Omega, \bC'',P, V_{\alpha}(\bC'))$ is not in general symmetric in the two questions unless $\alpha=0$ or $\alpha=1$. In the former case the relative depth reduces to the overlap $J(\Omega, (\bC';\bC''), P)$ which is symmetric and in the latter case the relative depth simply vanishes. Further, it can be seen from Fig.~\ref{f:relative-Y} that the relative depth can in fact be negative meaning that it is possible that the knowledge of an (imperfect) answer to a question may make another question more difficult. It would be interesting to establish general conditions under which relative depth is nonnegative. Another useful observation is that if for a pair of questions $\bC'$ and $\bC''$ question $\bC'$ is the more difficult one of the two then it appears that the inequality $Y(\Omega, \bC'', P, V_{\alpha}(\bC')) > Y(\Omega, \bC', P, V_{\alpha}(\bC''))$ holds for $0<\alpha < 1$ implying that a quasi-perfect answer to a more difficult question result in a higher reduction of difficulty of the other question. It would be of interest to see if this property holds in the general case or exceptions are possible.

\begin{figure}
\includegraphics[scale=0.7]{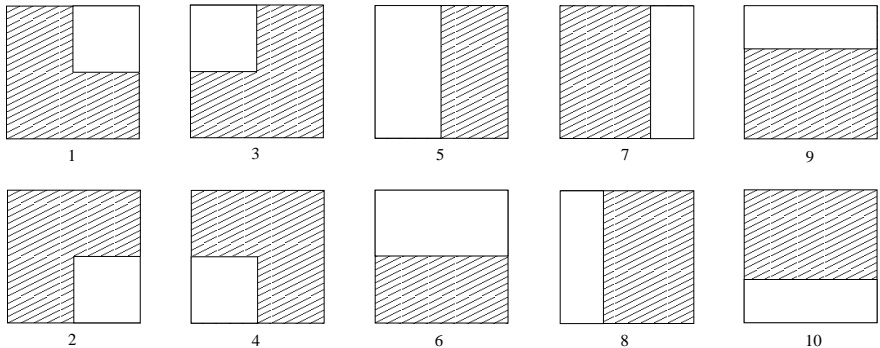}
\caption{\label{f:squaresubsets}Sample questions.}
\end{figure}

To illustrate the process of estimating the pseudotemperature $u(\o)$ and source model parameters, consider an example in which $\Omega=[0,1]^2\subset \bR^2$, and the measure $P$ is uniform continuous on $\Omega$. Consider the set of sample (complete) questions illustrated in Fig.~\ref{f:squaresubsets}.
Our goal is, given the error parameters $\alpha_i$ for quasi-perfect answer $V_{\alpha_i}(\bC_i)$ to question $\bC_i$, $i=1,\dotsc, 10$, estimate the function $u(\o)$ and the parameter(s) of the chosen information source model.

We adopt the modified linear source model and use formulation (\ref{eq:uest-mcm}) to estimate $u(\o)$, and parameters $Y_s$ and $b$ of the model. We do this for different values of error probabilities.

First consider data shown in Table~\ref{t:perffit}. In this and following tables, the first column contains the index $i$ of question $\bC_i$ from Fig.~\ref{f:squaresubsets}, the second column shows the corresponding error probability $\alpha_i$, and the last two columns contain the question difficulty $G(\Omega,\bC_i,P)$ and answer depth  $Y(\Omega,\bC_i,P,V_{\alpha_i}(\bC_i))$, respectively, obtained from the estimated values of $u(\o)$ and parameters of the source model. In the lower part of Table~\ref{t:perffit}, the resulting value of the objective of problem (\ref{eq:uest-mcm}) along with the estimated values of parameters $Y_s$ and $b$ are shown.

\begin{figure}
\includegraphics[scale=0.9]{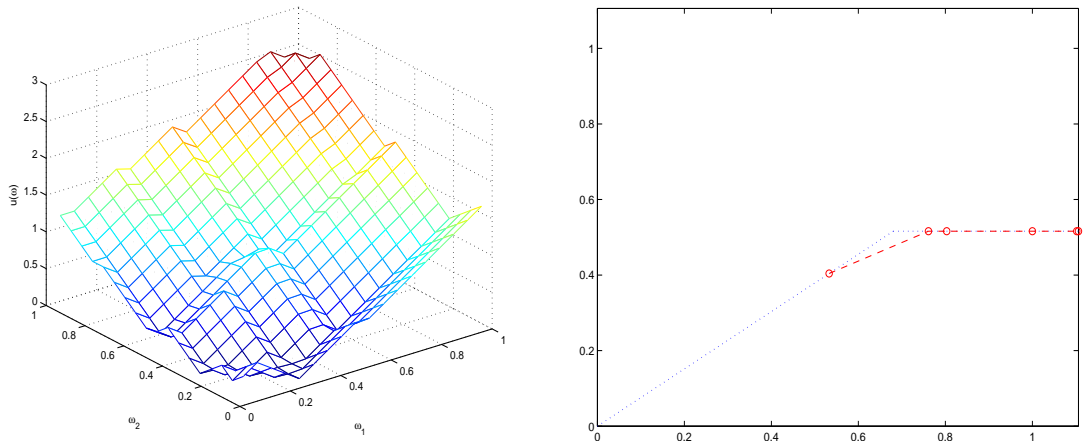}
\caption{\label{f:perffit}The estimated pseudotemperature (left) and the fitted values of difficulty and depth (right) for the data of Table~\ref{t:perffit}.}
\end{figure}

The error probability values shown in Table~\ref{t:perffit} result in a perfect fit ($z=0$) with the
estimated pseudotemperature function $u(\o)$ (shown in Fig.~\ref{f:perffit}). We can see that the resulting pseudotemperature function increases for the larger values of coordinates $\o_1$ and $\o_2$ on $\Omega$ reflecting the fact that, for instance $\alpha_1>\alpha_4$, implying that question $\bC_1$ has higher difficulty (larger value of pseudoenergy) than $\bC_4$ in spite of these two questions having same value of entropy. This means that the smaller measure subset in $\bC_1$ has to have higher pseudotemperature which we indeed see. It is also worth noting that questions $\bC_5$ and $\bC_6$ were answered with equal accuracy suggesting that these questions are of equal difficulty. This in fact is a necessary condition for a perfect fit within the ideal gas question difficulty model since in this model any complete question with all subsets of equal measure would have the same difficulty (pseudoenergy) regardless of the pseudotemperature function form.

\begin{table}
\caption{\label{t:perffit}Sample question error probabilities, fitted values of the difficulty and depth functions, and estimated model parameter values for the modified linear model when perfect fit is possible. {$\sum_{i=1}^{N_d} z_i = 0$; $ U=0.13$; $Y_s=0.52$; $b=0.76$.}}
{\begin{tabular*}{0.6\textwidth}%
     {@{\extracolsep{\fill}}cccc}
\hline
{$i$} & $\alpha_i$ & $G(\Omega,\bC_i,P)$ & $Y(\Omega,\bC_i,P,V_{\alpha_i}(\bC_i))$\\
\hline
 1 & 0.265 & 1.106 & 0.516 \\
2 & 0.143 & 0.803 & 0.516 \\
3 & 0.143 & 0.803 & 0.516 \\
4 & 0.077 & 0.533 & 0.404 \\
5 & 0.210 & 1.000 & 0.516 \\
6 & 0.210 & 1.000 & 0.516 \\
7 & 0.253 & 1.102 & 0.516 \\
8 & 0.116 & 0.761 & 0.516 \\
9 & 0.253 & 1.102 & 0.516 \\
10 & 0.116 & 0.761 & 0.516 \\
\hline
\end{tabular*}}
\end{table}

\begin{figure}
\includegraphics*[scale=0.9]{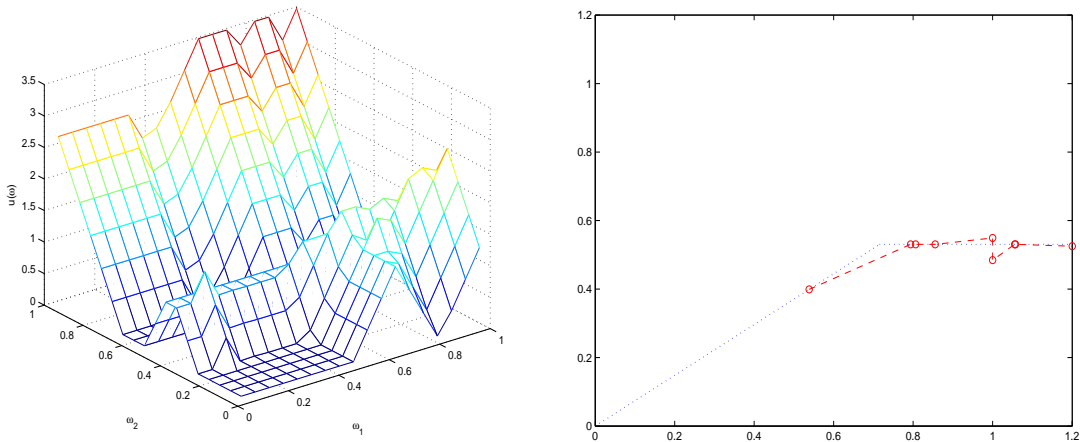}
\caption{\label{f:goodfit}The estimated pseudotemperature (left) and the fitted values of difficulty and depth (right) for the data of Table~\ref{t:goodfit}.}
\end{figure}

Consider now data shown in Table~\ref{t:goodfit}. The resulting pseudotemperature $u(\o)$ is shown in Fig.~\ref{f:goodfit}. We see that in this case the perfect fit could not be achieved by any pseudotemperature function, in particular because questions $\bC_5$ and $\bC_6$ were answered with slightly different accuracy whereas these two questions necessarily have equal pseudoenergy content (equal difficulty) within the ideal gas question difficulty model.

\begin{table}
\caption{\label{t:goodfit}Sample question error probabilities, fitted values of the difficulty and depth functions, and estimated model parameter values for the modified linear model when perfect fit is not possible, with small misfit. {$\sum_i^{N_d} z_i=0.07$; $U=0.43$; $Y_s=0.53$; $b=0.74$.}}
{\begin{tabular*}{0.6\textwidth}%
     {@{\extracolsep{\fill}}cccc}
\hline
{$i$} & $\alpha_i$ & $G(\Omega,\bC_i,P)$ & $Y(\Omega,\bC_i,P,V_{\alpha_i}(\bC_i))$\\
\hline
1 & 0.238 & 1.057 & 0.531 \\
2 & 0.157 & 0.856 & 0.531 \\
3 & 0.129 & 0.794 & 0.531 \\
4 & 0.084 & 0.538 & 0.399 \\
5 & 0.189 & 1.000 & 0.549 \\
6 & 0.230 & 1.000 & 0.484 \\
7 & 0.227 & 1.055 & 0.531 \\
8 & 0.127 & 0.806 & 0.531 \\
9 & 0.278 & 1.200 & 0.525 \\
10 & 0.127 & 0.806 & 0.531 \\
\hline
\end{tabular*}}
\end{table}

\begin{figure}
\includegraphics*[scale=0.9]{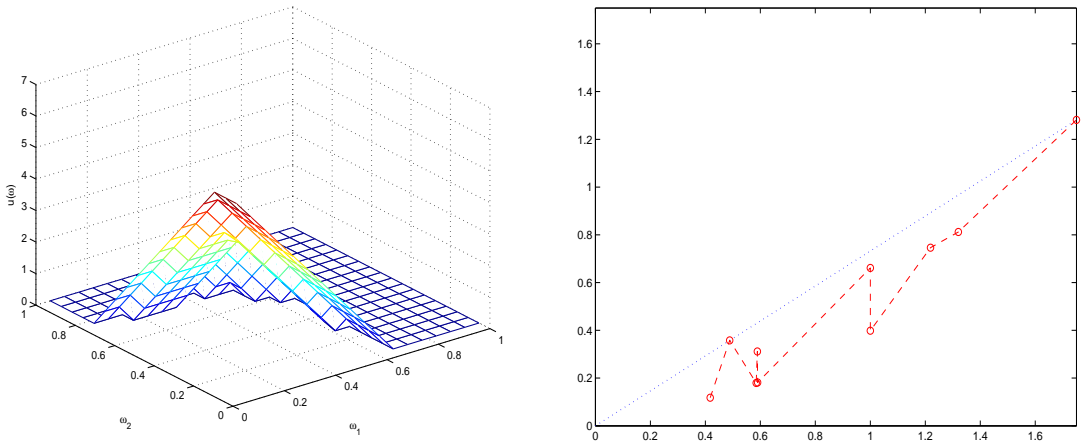}
\caption{\label{f:badfit}The estimated pseudotemperature (left) and the fitted values of difficulty and depth (right) for the data of Table~\ref{t:badfit}.}
\end{figure}

Now, consider the data shown in Table~\ref{t:badfit}. As can be seen from Fig.~\ref{f:badfit}, the fit that could be achieved to the ideal gas question difficulty model (with the linear modified information source model) is relatively (at least compared to the previous example) poor, possibly indicating that the ideal gas model may not be adequate in this case and that a different model (for example, anisotropic -- to be able to model different pseudoenergy content of questions $\bC_5$ and $\bC_6$) may be needed.

\begin{table}
\caption{\label{t:badfit}Sample question error probabilities, fitted values of the difficulty and depth functions, and estimated model parameter values for the modified linear model when perfect fit is not possible, with larger misfit. {$\sum_i^{N_d} z_i=1.51$; $U=0.56$; $Y_s=1.28$; $b=0.73$.}}
{\begin{tabular*}{0.6\textwidth}%
     {@{\extracolsep{\fill}}cccc}
\hline
{$i$} & $\alpha_i$ & $G(\Omega,\bC_i,P)$ & $Y(\Omega,\bC_i,P,V_{\alpha_i}(\bC_i))$\\
\hline
1 & 0.371 & 0.418 & 0.118 \\
2 & 0.086 & 0.488 & 0.358 \\
3 & 0.200 & 0.589 & 0.312 \\
4 & 0.107 & 1.750 & 1.281 \\
5 & 0.126 & 1.000 & 0.661 \\
6 & 0.293 & 1.000 & 0.399 \\
7 & 0.354 & 0.585 & 0.180 \\
8 & 0.162 & 1.320 & 0.812 \\
9 & 0.354 & 0.590 & 0.182 \\
10 & 0.162 & 1.219 & 0.746 \\
\hline
\end{tabular*}}
\end{table}

Let us now turn to comparing different sources. Suppose $\Omega=[0,1]$ with $P$ being a uniform continuous measure on $\Omega$. Let sample questions be as follows. $\bC_1=\{[0,1/2],(1/2,1]\}$, $\bC_2=\{[0,1/3],(1/3,1]\}$, $\bC_3=\{[0,2/3],(2/3,1]\}$, $\bC_4=\{[0,1/4],(1/4,1]\}$, $\bC_5=\{[0,3/4],(3/4,1]\}$. Let source 1 accuracy be described by error probabilities (assuming quasi-perfect answers as before) shown in Table~\ref{t:s1}. Then, using the modified capacity model and formulation (\ref{eq:uest-mcm}), we can estimate the pseudotemperature function $u(\cdot)$ and the model parameters $Y_s$ and $b$. The results -- as well as fitted values of the question difficulty and answer depth -- are shown in Table~\ref{t:s1}.

\begin{figure}
\includegraphics*[scale=0.5]{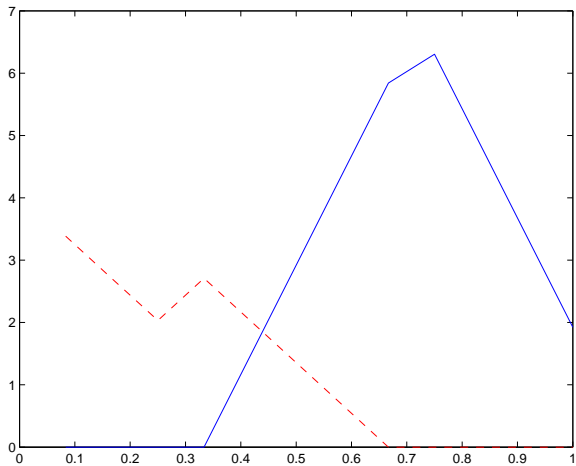}
\caption{\label{f:s12}Estimated pseudotemperature functions for information sources 1 and 2.}
\end{figure}

Table~\ref{t:s2} shows error probabilities achieved on the same set of sample questions by a different source 2, along with the resulting fitted values of difficulty and depth functions and the estimated model parameter values. Looking at Tables~\ref{t:s1} and~\ref{t:s2} we can see, for example, that source 1 shows better overall performance on all questions, but there exist questions (question 5, for instance) that appear to be easier for source 2. Indeed, the estimated pseudotemperature functions shown in Fig.~\ref{f:s12} (in the unit source capacity convention) clearly demonstrate that the overall pseudotemperature is significantly higher for source 2 thus making the majority of sample questions more difficult for it (which is reflected in higher error probabilities). On the other hand, while the pseudotemperature function for source 1 is (mostly) increasing on the interval $[0,1]$, it is a decreasing function on the same interval for source 2. In particular, there exist regions of $\Omega=[0,1]$ where the pseudotemperature for source 2 is lower than that for source 1. This means that some questions can be easier for source 2, question 5 from the sample set being an example.

\begin{table}
\caption{\label{t:s1}Sample question error probabilities, fitted values of the difficulty and depth functions, estimated model parameter values for the modified linear model, for information source 1. {$\sum_i^{N_d} z_i=0.09$; $U=0.54$; $Y_s=0.74$; $b=0.77$.}}
{\begin{tabular*}{0.6\textwidth}%
     {@{\extracolsep{\fill}}cccc}
\hline
{$i$} & $\alpha_i$ & $G(\Omega,\bC_i,P)$ & $Y(\Omega,\bC_i,P,V_{\alpha_i}(\bC_i))$\\
\hline
1 & 0.090 & 1.000 & 0.735 \\
2 & 0.070 & 0.678 & 0.525 \\
3 & 0.153 & 1.174 & 0.735 \\
4 & 0.070 & 0.528 & 0.408 \\
5 & 0.146 & 1.131 & 0.735 \\
\hline
\end{tabular*}}
\end{table}

\begin{table}
\caption{\label{t:s2}Sample question error probabilities, fitted values of the difficulty and depth functions, estimated model parameter values for the modified linear model, for information source 2.
{$\sum_i^{N_d} z_i=0.18$; $U=0.56$; $Y_s=0.39$; $b=0.74$.}}
{\begin{tabular*}{0.6\textwidth}%
     {@{\extracolsep{\fill}}cccc}
\hline
{$i$} & $\alpha_i$ & $G(\Omega,\bC_i,P)$ & $Y(\Omega,\bC_i,P,V_{\alpha_i}(\bC_i))$\\
\hline
1 & 0.300 & 0.933 & 0.386 \\
2 & 0.350 & 1.000 & 0.331 \\
3 & 0.170 & 0.415 & 0.229 \\
4 & 0.350 & 1.115 & 0.386 \\
5 & 0.080 & 0.585 & 0.434 \\
\hline
\end{tabular*}}
\end{table}

\section{\label{s:conclusion}Conclusion}
This article is devoted to developing quantitative framework for information exchange between a problem solving agent and an information source. The latter is assumed to be capable of providing answers to the agent's questions. While the companion article \cite{part1} is mostly concerned with questions, the main subject of the present article is answers and information source models. Questions can be characterized with the {\it question difficulty functional} which can be thought of as the amount of ``work'' the source would have to do in order to answer the particular question perfectly. The question difficulty functional is source-specific and reflects the {\it knowledge structure} of the source. The nature of geometric objects describing the knowledge structure is dictated by the symmetry exhibited by the latter. In particular, in an isotropic (linear) case, it is described by a scalar function while an anisotropic knowledge structure would likely be described -- as indicated by a preliminary investigation -- by a symmetric rank 2 tensor. The corresponding characterization of answers -- the {\it answer depth functional} -- is developed in the present article. It can be informally thought of as a measure of the amount of ``work'' the source is actually capable of doing in response to a particular question. The overall form of the answer depth functional reflects the source's knowledge structure and therefore largely parallels that of question difficulty. The value of answer depth is equal to that of question difficulty in case of a perfectly accurate answer and is less than question difficulty if the answer allows for errors.

Information source models describe the relationship between answer depth and question difficulty. It can be said that, while question difficulty reflects the source's {\it knowledge structure}, the information source model specifies the source's {\it knowledge strength} by relating the answer depth the source is capable of producing to the corresponding question difficulty. It is reasonable to expect that most sources would exhibit a sort of an upper bound on the achievable answer depth which could be identified with the source {\it capacity}. With regard to the latter, it is worth pointing out that this is not information capacity but rather {\it pseudoenergy} capacity which is the measure of maximum ``work'' the particular source is capable of.

The framework for describing information exchange between the agent and information sources is developed as part of a theory of the {\it Full Information Chain} which is anticipated to be an extension of the classical Information Theory in the direction of quantitative study of information {\it accuracy} and {\it relevance} attributes in addition to information {\it quantity} that is the main concern of Information Theory. The framework developed in \cite{part1} and the present article contains the basics of a theory of information acquisition, with information accuracy being the main attribute involved. The information usage link that deals with information relevance attribute is the subject of future publications. It should be noted that the information acquisition and usage appear to be fundamentally interconnected and have to be treated by a joint theory. Therefore any practical algorithms for optimal information acquisition will also be presented in future publications following a description of  the information usage link.





\begin{thebibliography}{46}%
\makeatletter
\providecommand \@ifxundefined [1]{%
 \@ifx{#1\undefined}
}%
\providecommand \@ifnum [1]{%
 \ifnum #1\expandafter \@firstoftwo
 \else \expandafter \@secondoftwo
 \fi
}%
\providecommand \@ifx [1]{%
 \ifx #1\expandafter \@firstoftwo
 \else \expandafter \@secondoftwo
 \fi
}%
\providecommand \natexlab [1]{#1}%
\providecommand \enquote  [1]{``#1''}%
\providecommand \bibnamefont  [1]{#1}%
\providecommand \bibfnamefont [1]{#1}%
\providecommand \citenamefont [1]{#1}%
\providecommand \href@noop [0]{\@secondoftwo}%
\providecommand \href [0]{\begingroup \@sanitize@url \@href}%
\providecommand \@href[1]{\@@startlink{#1}\@@href}%
\providecommand \@@href[1]{\endgroup#1\@@endlink}%
\providecommand \@sanitize@url [0]{\catcode `\\12\catcode `\$12\catcode
  `\&12\catcode `\#12\catcode `\^12\catcode `\_12\catcode `\%12\relax}%
\providecommand \@@startlink[1]{}%
\providecommand \@@endlink[0]{}%
\providecommand \url  [0]{\begingroup\@sanitize@url \@url }%
\providecommand \@url [1]{\endgroup\@href {#1}{\urlprefix }}%
\providecommand \urlprefix  [0]{URL }%
\providecommand \Eprint [0]{\href }%
\providecommand \doibase [0]{http://dx.doi.org/}%
\providecommand \selectlanguage [0]{\@gobble}%
\providecommand \bibinfo  [0]{\@secondoftwo}%
\providecommand \bibfield  [0]{\@secondoftwo}%
\providecommand \translation [1]{[#1]}%
\providecommand \BibitemOpen [0]{}%
\providecommand \bibitemStop [0]{}%
\providecommand \bibitemNoStop [0]{.\EOS\space}%
\providecommand \EOS [0]{\spacefactor3000\relax}%
\providecommand \BibitemShut  [1]{\csname bibitem#1\endcsname}%
\let\auto@bib@innerbib\@empty
\bibitem [{\citenamefont {Shannon}(1948)}]{SHANNON:1948}%
  \BibitemOpen
  \bibfield  {author} {\bibinfo {author} {\bibfnamefont {C.~E.}\ \bibnamefont
  {Shannon}},\ }\href@noop {} {\bibfield  {journal} {\bibinfo  {journal} {Bell
  Systems Technical Journal}\ }\textbf {\bibinfo {volume} {27}},\ \bibinfo
  {pages} {379} (\bibinfo {year} {1948})}\BibitemShut {NoStop}%
\bibitem [{\citenamefont {Perevalov}\ and\ \citenamefont
  {Grace}(2012)}]{part1}%
  \BibitemOpen
  \bibfield  {author} {\bibinfo {author} {\bibfnamefont {E.}~\bibnamefont
  {Perevalov}}\ and\ \bibinfo {author} {\bibfnamefont {D.}~\bibnamefont
  {Grace}},\ }\href@noop {} {\enquote {\bibinfo {title} {Towards the full
  information chain theory: question difficulty},}\ } (\bibinfo {year}
  {2012}),\ \bibinfo {note} {submitted to Physical Review E,
  arXiv:1212.2696[physics.data-an]}\BibitemShut {NoStop}%
\bibitem [{Note1()}]{Note1}%
  \BibitemOpen
  \bibinfo {note} {Note that this is not {\protect \it information} capacity
  like that of channel in classical Information Theory, but rather {\protect
  \it pseudoenergy} capacity.}\BibitemShut {Stop}%
\bibitem [{Note2()}]{Note2}%
  \BibitemOpen
  \bibinfo {note} {Finding this to be false for some questions would imply that
  the current description of the source's knowledge structure is not
  sufficiently accurate and that, for example, a more elaborate model is
  needed.}\BibitemShut {Stop}%
\bibitem [{\citenamefont {Viola}(1995)}]{viola1995}%
  \BibitemOpen
  \bibfield  {author} {\bibinfo {author} {\bibfnamefont {P.~A.}\ \bibnamefont
  {Viola}},\ }\href@noop {} {\emph {\bibinfo {title} {Alignment by maximization
  of mutual information}}},\ \bibinfo {type} {A.{I}. {T}echnical {R}eport}\
  \bibinfo {number} {1548}\ (\bibinfo  {institution} {Massachusetts Institute
  of Technology},\ \bibinfo {year} {1995})\BibitemShut {NoStop}%
\bibitem [{\citenamefont {Mokhov}\ and\ \citenamefont
  {Smirnov}(2006)}]{mokhov2006}%
  \BibitemOpen
  \bibfield  {author} {\bibinfo {author} {\bibfnamefont {I.~I.}\ \bibnamefont
  {Mokhov}}\ and\ \bibinfo {author} {\bibfnamefont {D.~A.}\ \bibnamefont
  {Smirnov}},\ }\href@noop {} {\bibfield  {journal} {\bibinfo  {journal}
  {Geophys. Res. Lett.}\ }\textbf {\bibinfo {volume} {33}} (\bibinfo {year}
  {2006})},\ \bibinfo {note} {l03708}\BibitemShut {NoStop}%
\bibitem [{\citenamefont {Verdes}(2005)}]{verdes2005}%
  \BibitemOpen
  \bibfield  {author} {\bibinfo {author} {\bibfnamefont {P.~F.}\ \bibnamefont
  {Verdes}},\ }\href@noop {} {\bibfield  {journal} {\bibinfo  {journal} {Phys.
  Rev. E}\ }\textbf {\bibinfo {volume} {72}} (\bibinfo {year} {2005})},\
  \bibinfo {note} {026222}\BibitemShut {NoStop}%
\bibitem [{\citenamefont {Katura}\ \emph {et~al.}(2006)\citenamefont {Katura},
  \citenamefont {Tanaka}, \citenamefont {Obata}, \citenamefont {Sato},\ and\
  \citenamefont {Maki}}]{katura2006}%
  \BibitemOpen
  \bibfield  {author} {\bibinfo {author} {\bibfnamefont {T.}~\bibnamefont
  {Katura}}, \bibinfo {author} {\bibfnamefont {N.}~\bibnamefont {Tanaka}},
  \bibinfo {author} {\bibfnamefont {A.}~\bibnamefont {Obata}}, \bibinfo
  {author} {\bibfnamefont {H.}~\bibnamefont {Sato}}, \ and\ \bibinfo {author}
  {\bibfnamefont {A.}~\bibnamefont {Maki}},\ }\href@noop {} {\bibfield
  {journal} {\bibinfo  {journal} {NeuroImage}\ }\textbf {\bibinfo {volume}
  {31}},\ \bibinfo {pages} {1592} (\bibinfo {year} {2006})}\BibitemShut
  {NoStop}%
\bibitem [{\citenamefont {Ch\'{a}vez}\ \emph {et~al.}(2003)\citenamefont
  {Ch\'{a}vez}, \citenamefont {Martinerie},\ and\ \citenamefont
  {Le~Van~Quyen}}]{chavez2003}%
  \BibitemOpen
  \bibfield  {author} {\bibinfo {author} {\bibfnamefont {M.}~\bibnamefont
  {Ch\'{a}vez}}, \bibinfo {author} {\bibfnamefont {J.}~\bibnamefont
  {Martinerie}}, \ and\ \bibinfo {author} {\bibfnamefont {M.}~\bibnamefont
  {Le~Van~Quyen}},\ }\href@noop {} {\bibfield  {journal} {\bibinfo  {journal}
  {J. of Neurosci. Methods}\ }\textbf {\bibinfo {volume} {124}},\ \bibinfo
  {pages} {113} (\bibinfo {year} {2003})}\BibitemShut {NoStop}%
\bibitem [{\citenamefont {Mc{C}ardle}(1985)}]{mccardle1985}%
  \BibitemOpen
  \bibfield  {author} {\bibinfo {author} {\bibfnamefont {K.~F.}\ \bibnamefont
  {Mc{C}ardle}},\ }\href@noop {} {\bibfield  {journal} {\bibinfo  {journal}
  {Management Sci.}\ }\textbf {\bibinfo {volume} {31}},\ \bibinfo {pages}
  {1372} (\bibinfo {year} {1985})}\BibitemShut {NoStop}%
\bibitem [{\citenamefont {Jensen}(1988)}]{jensen1988}%
  \BibitemOpen
  \bibfield  {author} {\bibinfo {author} {\bibfnamefont {R.}~\bibnamefont
  {Jensen}},\ }\href@noop {} {\bibfield  {journal} {\bibinfo  {journal}
  {Management Sci.}\ }\textbf {\bibinfo {volume} {34}},\ \bibinfo {pages} {230}
  (\bibinfo {year} {1988})}\BibitemShut {NoStop}%
\bibitem [{\citenamefont {Fisher}\ and\ \citenamefont
  {Raman}(1996)}]{fisher1996}%
  \BibitemOpen
  \bibfield  {author} {\bibinfo {author} {\bibfnamefont {M.~L.}\ \bibnamefont
  {Fisher}}\ and\ \bibinfo {author} {\bibfnamefont {A.}~\bibnamefont {Raman}},\
  }\href@noop {} {\bibfield  {journal} {\bibinfo  {journal} {Oper. Res.}\
  }\textbf {\bibinfo {volume} {44}},\ \bibinfo {pages} {87} (\bibinfo {year}
  {1996})}\BibitemShut {NoStop}%
\bibitem [{\citenamefont {Kornish}\ and\ \citenamefont
  {Keeney}(2008)}]{kornish2008}%
  \BibitemOpen
  \bibfield  {author} {\bibinfo {author} {\bibfnamefont {L.~J.}\ \bibnamefont
  {Kornish}}\ and\ \bibinfo {author} {\bibfnamefont {R.~L.}\ \bibnamefont
  {Keeney}},\ }\href@noop {} {\bibfield  {journal} {\bibinfo  {journal} {Oper.
  Res.}\ }\textbf {\bibinfo {volume} {56}},\ \bibinfo {pages} {527} (\bibinfo
  {year} {2008})}\BibitemShut {NoStop}%
\bibitem [{\citenamefont {Fischer}\ \emph {et~al.}(1996)\citenamefont
  {Fischer}, \citenamefont {Arnold},\ and\ \citenamefont
  {Gibbs}}]{fischer1996}%
  \BibitemOpen
  \bibfield  {author} {\bibinfo {author} {\bibfnamefont {A.~J.}\ \bibnamefont
  {Fischer}}, \bibinfo {author} {\bibfnamefont {A.~J.}\ \bibnamefont {Arnold}},
  \ and\ \bibinfo {author} {\bibfnamefont {M.}~\bibnamefont {Gibbs}},\
  }\href@noop {} {\bibfield  {journal} {\bibinfo  {journal} {Amer. J. Agr.
  Econ.}\ }\textbf {\bibinfo {volume} {78}},\ \bibinfo {pages} {1073} (\bibinfo
  {year} {1996})}\BibitemShut {NoStop}%
\bibitem [{\citenamefont {Ellison}\ and\ \citenamefont
  {Fudenberg}(1993)}]{ellison1993}%
  \BibitemOpen
  \bibfield  {author} {\bibinfo {author} {\bibfnamefont {G.}~\bibnamefont
  {Ellison}}\ and\ \bibinfo {author} {\bibfnamefont {D.}~\bibnamefont
  {Fudenberg}},\ }\href@noop {} {\bibfield  {journal} {\bibinfo  {journal} {J.
  Political Econom.}\ }\textbf {\bibinfo {volume} {101}},\ \bibinfo {pages}
  {612} (\bibinfo {year} {1993})}\BibitemShut {NoStop}%
\bibitem [{\citenamefont {French}(1985)}]{french1985}%
  \BibitemOpen
  \bibfield  {author} {\bibinfo {author} {\bibfnamefont {S.}~\bibnamefont
  {French}},\ }\href@noop {} {\bibfield  {journal} {\bibinfo  {journal}
  {Bayesian Statist.}\ }\textbf {\bibinfo {volume} {2}},\ \bibinfo {pages}
  {183} (\bibinfo {year} {1985})}\BibitemShut {NoStop}%
\bibitem [{\citenamefont {Genest}\ and\ \citenamefont
  {Zidek}(1986)}]{genest1986}%
  \BibitemOpen
  \bibfield  {author} {\bibinfo {author} {\bibfnamefont {C.}~\bibnamefont
  {Genest}}\ and\ \bibinfo {author} {\bibfnamefont {J.~V.}\ \bibnamefont
  {Zidek}},\ }\href@noop {} {\bibfield  {journal} {\bibinfo  {journal}
  {Statist. Sci.}\ }\textbf {\bibinfo {volume} {1}},\ \bibinfo {pages} {114}
  (\bibinfo {year} {1986})}\BibitemShut {NoStop}%
\bibitem [{\citenamefont {Clemen}(1987)}]{clemen1987}%
  \BibitemOpen
  \bibfield  {author} {\bibinfo {author} {\bibfnamefont {R.}~\bibnamefont
  {Clemen}},\ }\href@noop {} {\bibfield  {journal} {\bibinfo  {journal}
  {Management Sci.}\ }\textbf {\bibinfo {volume} {33}},\ \bibinfo {pages} {373}
  (\bibinfo {year} {1987})}\BibitemShut {NoStop}%
\bibitem [{\citenamefont {Clemen}\ and\ \citenamefont
  {Winkler}(1999)}]{clemen1999}%
  \BibitemOpen
  \bibfield  {author} {\bibinfo {author} {\bibfnamefont {R.}~\bibnamefont
  {Clemen}}\ and\ \bibinfo {author} {\bibfnamefont {R.}~\bibnamefont
  {Winkler}},\ }\href@noop {} {\bibfield  {journal} {\bibinfo  {journal} {Risk
  Anal.}\ }\textbf {\bibinfo {volume} {19}},\ \bibinfo {pages} {187} (\bibinfo
  {year} {1999})}\BibitemShut {NoStop}%
\bibitem [{\citenamefont {Predd}\ \emph {et~al.}(2008)\citenamefont {Predd},
  \citenamefont {Osherson}, \citenamefont {Kulkarni},\ and\ \citenamefont
  {Poor}}]{predd2008}%
  \BibitemOpen
  \bibfield  {author} {\bibinfo {author} {\bibfnamefont {J.~B.}\ \bibnamefont
  {Predd}}, \bibinfo {author} {\bibfnamefont {D.~N.}\ \bibnamefont {Osherson}},
  \bibinfo {author} {\bibfnamefont {S.~R.}\ \bibnamefont {Kulkarni}}, \ and\
  \bibinfo {author} {\bibfnamefont {H.~V.}\ \bibnamefont {Poor}},\ }\href@noop
  {} {\bibfield  {journal} {\bibinfo  {journal} {Decision Anal.}\ }\textbf
  {\bibinfo {volume} {5}},\ \bibinfo {pages} {177} (\bibinfo {year}
  {2008})}\BibitemShut {NoStop}%
\bibitem [{\citenamefont {Bordley}(2009)}]{bordley2009}%
  \BibitemOpen
  \bibfield  {author} {\bibinfo {author} {\bibfnamefont {R.~F.}\ \bibnamefont
  {Bordley}},\ }\href@noop {} {\bibfield  {journal} {\bibinfo  {journal}
  {Decision Anal.}\ }\textbf {\bibinfo {volume} {6}},\ \bibinfo {pages} {38}
  (\bibinfo {year} {2009})}\BibitemShut {NoStop}%
\bibitem [{\citenamefont {Bordley}(2011)}]{bordley2011}%
  \BibitemOpen
  \bibfield  {author} {\bibinfo {author} {\bibfnamefont {R.~F.}\ \bibnamefont
  {Bordley}},\ }\href@noop {} {\bibfield  {journal} {\bibinfo  {journal}
  {Decision Anal.}\ }\textbf {\bibinfo {volume} {8}},\ \bibinfo {pages} {117}
  (\bibinfo {year} {2011})}\BibitemShut {NoStop}%
\bibitem [{\citenamefont {Faddeev}(1956)}]{faddeev1956}%
  \BibitemOpen
  \bibfield  {author} {\bibinfo {author} {\bibfnamefont {D.~K.}\ \bibnamefont
  {Faddeev}},\ }\href@noop {} {\bibfield  {journal} {\bibinfo  {journal}
  {Uspekhi Mat. Nauk}\ }\textbf {\bibinfo {volume} {11}},\ \bibinfo {pages}
  {227} (\bibinfo {year} {1956})}\BibitemShut {NoStop}%
\bibitem [{\citenamefont {R{\'e}nyi}(1961)}]{RENYI:1961}%
  \BibitemOpen
  \bibfield  {author} {\bibinfo {author} {\bibfnamefont {A.}~\bibnamefont
  {R{\'e}nyi}},\ }in\ \href@noop {} {\emph {\bibinfo {booktitle} {Proc. 4th
  {B}erkeley {S}ympos. {M}ath. {S}tatist. and {P}rob., {V}ol. {I}}}}\ (\bibinfo
   {publisher} {Univ. California Press},\ \bibinfo {address} {Berkeley,
  Calif.},\ \bibinfo {year} {1961})\ pp.\ \bibinfo {pages}
  {547--561}\BibitemShut {NoStop}%
\bibitem [{\citenamefont {Havrda}\ and\ \citenamefont
  {Charvat}(1967)}]{havrda1967}%
  \BibitemOpen
  \bibfield  {author} {\bibinfo {author} {\bibfnamefont {J.~H.}\ \bibnamefont
  {Havrda}}\ and\ \bibinfo {author} {\bibfnamefont {F.}~\bibnamefont
  {Charvat}},\ }\href@noop {} {\bibfield  {journal} {\bibinfo  {journal}
  {Kybernetika}\ }\textbf {\bibinfo {volume} {3}},\ \bibinfo {pages} {30}
  (\bibinfo {year} {1967})}\BibitemShut {NoStop}%
\bibitem [{\citenamefont {Simovici}\ and\ \citenamefont
  {Jaroszewicz}(2002)}]{simovici2002}%
  \BibitemOpen
  \bibfield  {author} {\bibinfo {author} {\bibfnamefont {D.~A.}\ \bibnamefont
  {Simovici}}\ and\ \bibinfo {author} {\bibfnamefont {S.}~\bibnamefont
  {Jaroszewicz}},\ }\href@noop {} {\bibfield  {journal} {\bibinfo  {journal}
  {IEEE Trans. Inf. Theory}\ }\textbf {\bibinfo {volume} {48}},\ \bibinfo
  {pages} {2138} (\bibinfo {year} {2002})}\BibitemShut {NoStop}%
\bibitem [{\citenamefont {Jaroszewicz}\ and\ \citenamefont
  {Simovici}(1999)}]{simovici1999}%
  \BibitemOpen
  \bibfield  {author} {\bibinfo {author} {\bibfnamefont {S.}~\bibnamefont
  {Jaroszewicz}}\ and\ \bibinfo {author} {\bibfnamefont {D.~A.}\ \bibnamefont
  {Simovici}},\ }in\ \href@noop {} {\emph {\bibinfo {booktitle} {Proc. 29th
  ISMVL, Freiburg, Germany}}}\ (\bibinfo {year} {1999})\ pp.\ \bibinfo {pages}
  {24--31}\BibitemShut {NoStop}%
\bibitem [{\citenamefont {Caticha}(2012{\natexlab{a}})}]{caticha-rev}%
  \BibitemOpen
  \bibfield  {author} {\bibinfo {author} {\bibfnamefont {A.}~\bibnamefont
  {Caticha}},\ }\href@noop {} {\emph {\bibinfo {title} {Entropic Inference and
  the Foundations of Physics}}}\ (\bibinfo  {publisher} {11th Brazilian Meeting
  on Bayesian Statistics},\ \bibinfo {address} {S\~ao Paolo, Brazil},\ \bibinfo
  {year} {2012})\BibitemShut {NoStop}%
\bibitem [{\citenamefont {Jaynes}(1957{\natexlab{a}})}]{JAYNES:1957a}%
  \BibitemOpen
  \bibfield  {author} {\bibinfo {author} {\bibfnamefont {E.~T.}\ \bibnamefont
  {Jaynes}},\ }\href@noop {} {\bibfield  {journal} {\bibinfo  {journal} {Phys.
  Rev.}\ }\textbf {\bibinfo {volume} {106}},\ \bibinfo {pages} {620} (\bibinfo
  {year} {1957}{\natexlab{a}})}\BibitemShut {NoStop}%
\bibitem [{\citenamefont {Jaynes}(1957{\natexlab{b}})}]{JAYNES:1957b}%
  \BibitemOpen
  \bibfield  {author} {\bibinfo {author} {\bibfnamefont {E.~T.}\ \bibnamefont
  {Jaynes}},\ }\href@noop {} {\bibfield  {journal} {\bibinfo  {journal} {Phys.
  Rev.}\ }\textbf {\bibinfo {volume} {108}},\ \bibinfo {pages} {171} (\bibinfo
  {year} {1957}{\natexlab{b}})}\BibitemShut {NoStop}%
\bibitem [{\citenamefont {Caticha}\ and\ \citenamefont
  {Cafaro}(2007)}]{caticha07}%
  \BibitemOpen
  \bibfield  {author} {\bibinfo {author} {\bibfnamefont {A.}~\bibnamefont
  {Caticha}}\ and\ \bibinfo {author} {\bibfnamefont {C.}~\bibnamefont
  {Cafaro}},\ }in\ \href@noop {} {\emph {\bibinfo {booktitle} {Bayesian
  Inference and Maximum Entropy Methods in Science and Engineering}}},\
  \bibinfo {series} {AIP Conf. Proc.}, Vol.\ \bibinfo {volume} {954},\ \bibinfo
  {editor} {edited by\ \bibinfo {editor} {\bibfnamefont {K.}~\bibnamefont
  {Knuth}} \emph {et~al.}}\ (\bibinfo {year} {2007})\ p.\ \bibinfo {pages}
  {165}\BibitemShut {NoStop}%
\bibitem [{\citenamefont {Caticha}(2011)}]{caticha11}%
  \BibitemOpen
  \bibfield  {author} {\bibinfo {author} {\bibfnamefont {A.}~\bibnamefont
  {Caticha}},\ }\href@noop {} {\bibfield  {journal} {\bibinfo  {journal} {J.
  Phys.}\ }\textbf {\bibinfo {volume} {A 44}},\ \bibinfo {pages} {225303}
  (\bibinfo {year} {2011})}\BibitemShut {NoStop}%
\bibitem [{\citenamefont {Caticha}(2012{\natexlab{b}})}]{caticha12}%
  \BibitemOpen
  \bibfield  {author} {\bibinfo {author} {\bibfnamefont {A.}~\bibnamefont
  {Caticha}},\ }in\ \href@noop {} {\emph {\bibinfo {booktitle} {MaxEnt 2012,
  The 32nd International Workshop on Bayesian Inference and Maximum Entropy
  Methods in Science and Engineering}}}\ (\bibinfo {year} {2012})\ \bibinfo
  {note} {arXiv:1212.6946}\BibitemShut {NoStop}%
\bibitem [{\citenamefont {Cox}(1946)}]{cox1946}%
  \BibitemOpen
  \bibfield  {author} {\bibinfo {author} {\bibfnamefont {R.~T.}\ \bibnamefont
  {Cox}},\ }\href@noop {} {\bibfield  {journal} {\bibinfo  {journal} {Am. J.
  Phys.}\ }\textbf {\bibinfo {volume} {14}},\ \bibinfo {pages} {1} (\bibinfo
  {year} {1946})}\BibitemShut {NoStop}%
\bibitem [{\citenamefont {Cox}(1961)}]{cox1961}%
  \BibitemOpen
  \bibfield  {author} {\bibinfo {author} {\bibfnamefont {R.~T.}\ \bibnamefont
  {Cox}},\ }\href@noop {} {\emph {\bibinfo {title} {The Algebra of Probable
  Inference}}}\ (\bibinfo  {publisher} {Johns Hopkins Press},\ \bibinfo
  {address} {Baltimore},\ \bibinfo {year} {1961})\BibitemShut {NoStop}%
\bibitem [{\citenamefont {Cox}(1979)}]{cox1979}%
  \BibitemOpen
  \bibfield  {author} {\bibinfo {author} {\bibfnamefont {R.~T.}\ \bibnamefont
  {Cox}},\ }in\ \href@noop {} {\emph {\bibinfo {booktitle} {The Maximum Entropy
  Formalism}}},\ \bibinfo {editor} {edited by\ \bibinfo {editor} {\bibfnamefont
  {R.}~\bibnamefont {Levine}}\ and\ \bibinfo {editor} {\bibfnamefont
  {M.}~\bibnamefont {Tribus}}}\ (\bibinfo  {publisher} {MIT Press},\ \bibinfo
  {address} {Cambridge, MA},\ \bibinfo {year} {1979})\ \bibinfo {note} {pp.
  119-167}\BibitemShut {NoStop}%
\bibitem [{\citenamefont {Knuth}(2005)}]{knuth05}%
  \BibitemOpen
  \bibfield  {author} {\bibinfo {author} {\bibfnamefont {K.~H.}\ \bibnamefont
  {Knuth}},\ }\href@noop {} {\bibfield  {journal} {\bibinfo  {journal}
  {Neurocomputing}\ }\textbf {\bibinfo {volume} {67}},\ \bibinfo {pages}
  {245–274} (\bibinfo {year} {2005})}\BibitemShut {NoStop}%
\bibitem [{\citenamefont {Knuth}(2007)}]{knuth07}%
  \BibitemOpen
  \bibfield  {author} {\bibinfo {author} {\bibfnamefont {K.~H.}\ \bibnamefont
  {Knuth}},\ }in\ \href@noop {} {\emph {\bibinfo {booktitle} {Bayesian
  Inference and Maximum Entropy Methods in Science and Engineering}}},\
  \bibinfo {series} {AIP Conf. Proc.}, Vol.\ \bibinfo {volume} {954},\ \bibinfo
  {editor} {edited by\ \bibinfo {editor} {\bibfnamefont {K.}~\bibnamefont
  {Knuth}} \emph {et~al.}}\ (\bibinfo {year} {2007})\ pp.\ \bibinfo {pages}
  {23--36}\BibitemShut {NoStop}%
\bibitem [{\citenamefont {Knuth}(2008)}]{knuth08}%
  \BibitemOpen
  \bibfield  {author} {\bibinfo {author} {\bibfnamefont {K.~H.}\ \bibnamefont
  {Knuth}},\ }in\ \href@noop {} {\emph {\bibinfo {booktitle} {Bayesian
  Inference and Maximum Entropy Methods in Science and Engineering, Sao Paolo,
  Brazil}}}\ (\bibinfo {year} {2008})\ pp.\ \bibinfo {pages}
  {24--31}\BibitemShut {NoStop}%
\bibitem [{\citenamefont {Knuth}\ and\ \citenamefont
  {Bahreyni}(2010)}]{knuth-sr}%
  \BibitemOpen
  \bibfield  {author} {\bibinfo {author} {\bibfnamefont {K.~H.}\ \bibnamefont
  {Knuth}}\ and\ \bibinfo {author} {\bibfnamefont {N.}~\bibnamefont
  {Bahreyni}},\ }\href@noop {} {\enquote {\bibinfo {title} {A derivation of
  special relativity from causal sets},}\ }\bibinfo {howpublished}
  {arXiv:1005.4172v1[math-ph]} (\bibinfo {year} {2010})\BibitemShut {NoStop}%
\bibitem [{Note3()}]{Note3}%
  \BibitemOpen
  \bibinfo {note} {It should be noted at this point that the role of
  information source can be played by conscious agents (human experts) and
  various data sources alike. In the latter case, additional care has to be
  taken interpreting questions and answers but the overall construction still
  applies. This theme will be developed further in future
  publications.}\BibitemShut {Stop}%
\bibitem [{\citenamefont {Caticha}(2004)}]{caticha04}%
  \BibitemOpen
  \bibfield  {author} {\bibinfo {author} {\bibfnamefont {A.}~\bibnamefont
  {Caticha}},\ }in\ \href@noop {} {\emph {\bibinfo {booktitle} {Bayesian
  Inference and Maximum Entropy Methods in Science and Engineering}}},\
  \bibinfo {series} {AIP Conf. Proc.}, Vol.\ \bibinfo {volume} {735},\ \bibinfo
  {editor} {edited by\ \bibinfo {editor} {\bibfnamefont {R.}~\bibnamefont
  {Fischer}} \emph {et~al.}}\ (\bibinfo {year} {2004})\ p.\ \bibinfo {pages}
  {429}\BibitemShut {NoStop}%
\bibitem [{Note4()}]{Note4}%
  \BibitemOpen
  \bibinfo {note} {It is possible that other questions will play a role in
  future developments.}\BibitemShut {Stop}%
\bibitem [{Note5()}]{Note5}%
  \BibitemOpen
  \bibinfo {note} {For instance, in \cite {caticha04}, an example of an ideal
  question is given: {\protect \it ``What color is Napoleon's white horse?''}.
  In our interpretation, an ideal question corresponding to the element
  {\protect \it ``White''} of the base space $\Omega $ consisting of all
  possible horse colors would not be phrased this way. Rather, the verbalized
  question will simply be {\protect \it ``What color is Napoleon's horse?''}
  and the particular ideal question will be implicitly asked if and only if
  Napoleon's horse is indeed white. Obviously, the latter fact (that the ideal
  question being asked corresponds to the element {\protect \it ``White''}) can
  only be known to someone having certain knowledge of the true color of
  Napoleon's horse.}\BibitemShut {Stop}%
\bibitem [{Note6()}]{Note6}%
  \BibitemOpen
  \bibinfo {note} {Here, just like in \cite {part1}, we use the term ``correct
  answer'' for a complete question $\protect \mathbf {C}$ to denote the subset
  $C\in \protect \mathbf {C}$ such that $\omega \in C$ for the given instance
  of the question. Using this terminology, a perfect answer to a complete
  question is simply a message that identifies the correct answer with
  certainty. Any ideal question has a single correct answer and is fully
  described by the latter.}\BibitemShut {Stop}%
\bibitem [{Note7()}]{Note7}%
  \BibitemOpen
  \bibinfo {note} {For the sake of simplicity, we assume that the answers of
  the source are quasi-perfect with the corresponding (estimated) error
  probabilities being equal to $\alpha _1, \protect \dotsc , \alpha _K$,
  respectively.}\BibitemShut {Stop}%
\end{thebibliography}


%

\end{document}